# Superabsorption in an organic microcavity: towards a quantum battery

James Q. Quach[1]*, Kirsty E. McGhee[3], Lucia Ganzer[2], Dominic M. Rouse[4], Brendon W. Lovett[4], Erik M. Gauger[5], Jonathan Keeling[4], Giulio Cerullo[2], David G. Lidzey[3], Tersilla Virgili[2]*

1. Institute for Photonics and Advanced Sensing and School of Chemistry and Physics, The University of Adelaide, South Australia 5005, Australia

2. Istituto di Fotonica e Nanotecnologia – CNR, IFN - Dipartimento di Fisica, Politecnico di Milano, Piazza Leonardo da Vinci 32, 20133 Milano, Italy

3. Department of Physics and Astronomy, University of Sheffield, Hicks Building, Hounsfield Road, Sheffield S3 7RH, U.K.

4. SUPA, School of Physics and Astronomy, University of St Andrews, St Andrews, KY16 9SS, UK

5. SUPA, Institute of Photonics and Quantum Sciences, Heriot-Watt University, EH14 4AS, UK

*Corresponding authors: quach.james@gmail.com, tersilla.virgili@polimi.it

**The rate at which matter emits or absorbs light can be modified by its environment, as dramatically exemplified by the widely-studied phenomenon of superradiance. The reverse process, superabsorption, is harder to demonstrate due to the challenges of probing ultrafast processes, and has only been seen for small numbers of atoms. Its central idea—superextensive scaling of absorption meaning larger systems absorb faster—is also the key idea underpinning quantum batteries. Here we implement experimentally a paradigmatic model of a quantum battery, constructed of a microcavity enclosing a molecular dye. Ultrafast optical spectroscopy allows us**

to observe charging dynamics at femtosecond resolution to demonstrate superextensive charging rates and storage capacity, in agreement with our theoretical modelling. We find that decoherence plays an important role in stabilising energy storage. Our work opens new opportunities for harnessing collective effects in light-matter coupling for nanoscale energy capture, storage, and transport technologies.

## Introduction

The properties of physical systems can typically be categorised as intensive (i.e. they are independent of the system size, such as density) or extensive (i.e. they grow in proportion to system size, such as mass). However, in some cases, co-operative effects can lead to super-extensive scaling. A well-studied example of this is superradiant emission[1]. In its original form, this describes emission from an ensemble of $N$ emitters into free space. Constructive interference in the emission process means that the time for emission scales as $1/N$, so that peak emission power is superextensive, scaling as $N^2$. Such behaviour has been demonstrated on a number of platforms (low pressure gases[2, 3], quantum wells[4, 5] and dots[6], J aggregates[7], Bose-Einstein condensates[8], trapped atoms[9], nitrogen-vacancy centres[10]). A less-studied example is superabsorption[11], describing the $N$-dependent enhancement of absorption of radiation by an ensemble of $N$ two-level systems (TLSs). Only very recently has this been demonstrated for a small number of atoms[12]. In principle, superabsorption could have important implications for energy storage and capture technologies, particularly if realised in platforms compatible with energy harvesting, such as organic photovoltaic devices. However, there are challenges in engineering the precise environment in which such behaviour can occur, and in monitoring the ultrashort charging time scales. In this article we show how these can be overcome, by combining organic microcavity fabrication with ultrafast pump-probe spectroscopy.

Superextensive scaling of energy absorption is also a key property of quantum batteries (QB). These represent a new class of energy storage devices that operate on distinctly quantum mechanical principles. In particular, they are driven either by quantum entanglement, that reduces the number of traversed states in the Hilbert space compared to (classical) separable states alone[13-21], or by cooperative behaviour that increases the effective quantum coupling between battery and source[22-24]. Such effects mean that QBs exhibit a charging time that is inversely related to the battery capacity. This leads to the intriguing idea that the charging power of QBs is superextensive; that is, it increases faster than the size of the battery. For a QB consisting of a collection of $N$ identical quantum systems, superextensive charging rate density (charging rate per subsystem) that scales as $N$ or $\sqrt{N}$ in the thermodynamic limit[20], has been predicted.

Here we experimentally realise a paradigmatic model proposed as a Dicke QB[24], which displays superextensive scaling of energy absorption, using an organic semiconductor as an ensemble of TLSs coupled to a confined optical mode in a microcavity. We also demonstrate how dissipation plays a crucial role; in a closed system, the coherent effects that lead to fast charging can also lead to subsequent fast discharging. As such, stabilisation of stored energy remains an open question: proposed stabilisation methods include continuous measurements[25], dark states[21], and novel energy trapping mechanisms[26, 27]. In our open noisy system, dephasing causes transitions between the optically active bright mode, and inactive dark modes. This suppresses emission into the cavity mode, so that we have fast absorption of energy but slow decay, allowing retention of the stored energy until it can be used.

**Results and Discussion**

**Device structure**

The structures fabricated consist of a thin (active) layer of a low-mass molecular semiconductor dispersed into a polymer matrix that is deposited by spin-coating and positioned between two dielectric mirrors, forming a microcavity as illustrated schematically in Fig. 1a (see Methods section for fabrication details). Organic semiconductors are particularly promising for many applications as the high oscillator strength and binding energy of molecular excitons means that light can be absorbed efficiently and excitons can exist at room temperature(*28*). The organic semiconductor used in this study was the dye Lumogen-F Orange (LFO), whose chemical structure is shown in Fig. 1b. The normalised absorption and photoluminescence spectra for LFO dispersed at 1% concentration by mass in a polystyrene (PS) matrix are shown in Fig. 1b. By diluting the LFO, we reduce intermolecular interactions that lead to emission quenching, producing a high photoluminescence quantum yield of around 60% at low concentration (see Supplemental Fig. S1). The absorption peak at 526 nm and the emission peak at 534 nm correspond to the 0-0 transition, i.e. an electronic transition from and to the lowest vibrational state. Operating around the 0-0 transition, the LFO molecules can reasonably be considered as a TLS. We prepared samples with 0.5%, 1%, 5%, and 10% concentrations, as these are representative of the optimal operating regimes - further increases in concentration lead to quenching, and signals from lower concentrations are indiscernible from noise. The absorption and photoluminescence spectra for the 0.5%, 5% and 10% concentrations are given in Supplemental Fig. S2.

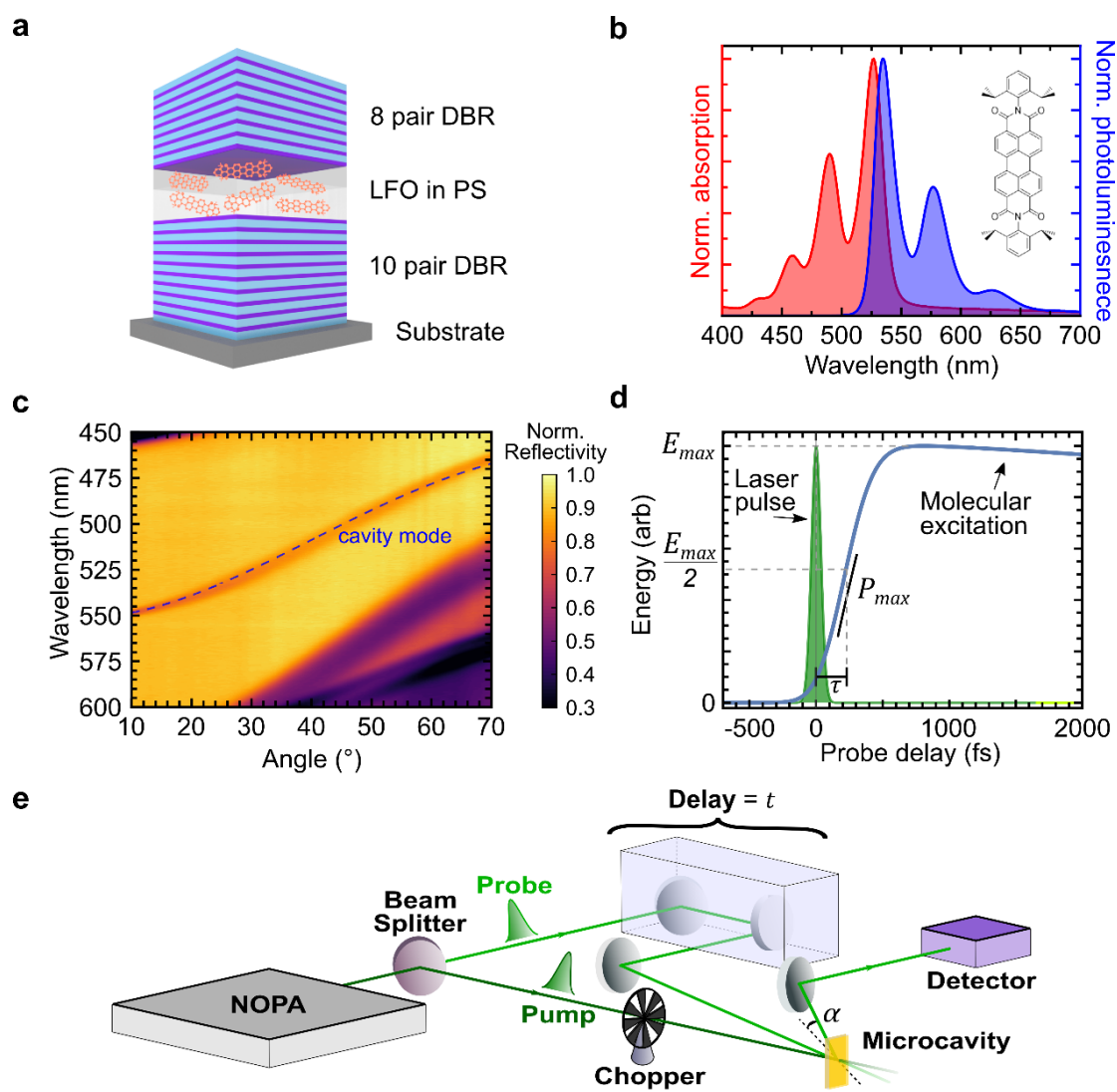

**Fig. 1 Schematics of the LFO microcavity and experimental setup.** (**a**) Microcavity consisting of Lumogen-F Orange (LFO) dispersed in a polystyrene (PS) matrix between distributed Bragg reflectors (DBRs). (**b**) Normalised absorption (red) and photoluminescence (blue) spectra for 1% concentration LFO film, with the molecular structure shown in the inset. We operate near peak absorption/photoluminescence. (**c**) Angle-dependent reflectivity of the 1% cavity, with a fit for the cavity mode shown by the blue dashed line. (**d**) A laser pump pulse excites the LFO molecules. The energetics of the molecules are then measured with probe pulses delayed by time $t$, from which we can ascertain the peak energy density ($E_{max}$), rise time ($\tau$), and peak charging power ($P_{max}$). (**e**) The experimental setup for ultrafast transient reflectivity measurements. The output of a non-collinear optical parametric amplifier (NOPA) is split to generate pump (dark green) and probe (light green) pulses. A mechanical chopper is used to modulate the pump pulse to produce alternating pump-probe and probe-only pulses.

The optical microcavities fabricated support cavity modes whose energy is determined by the optical-thickness of the LFO layer and the penetration of the optical field into the cavity mirrors(*29*). The confined photon field drives coherent interactions with the molecules, which underpin the collective effects that drive superabsorption. The LFO concentration dictates the operating coupling regime, with the 0.5% and 1% LFO cavities operating in the weak coupling regime, the 5% in the intermediate coupling regime, and the 10% in the strong coupling regime (see Supplemental Fig. S2 and discussion in Methods).

**Experimental setup**

Charging and energy storage dynamics were measured using ultrafast transient-absorption spectroscopy(*30*), allowing femtosecond charging times to be measured. In this technique, we excite the microcavity with a pump pulse, and then measure the evolution of stored energy (i.e. corresponding to the number of excited molecules) with a second probe pulse, delayed by time $t$ (Fig. 1d). The probe pulse is transmitted through the top distributed Bragg reflector (DBR) of the cavity, and the reflection from the bottom DBR is measured. The differential reflectivity induced by the pump-pulse is given by

$$\frac{\Delta R}{R}(t) = \frac{R_{ON}(t) - R_{OFF}}{R_{OFF}}, \tag{3}$$

where $R_{ON}$ ($R_{OFF}$) is the probe reflectivity with (without) the pump excitation. Note that control films (active layers without the microcavities) are measured under differential transmittivity $\Delta T / T$. The control films will allow us to identify the underlying photophysics of the molecules.

In our experimental setup (shown schematically in Fig. 1e), transient-absorption measurements were performed in a degenerate, almost collinear configuration. Pump and probe pulses were generated by a broadband non-collinear optical parametric amplifier (NOPA)(*31*) and spanned the wavelength range 500 to 620 nm with a nearly transform-limited sub-20-fs duration (further details in Methods). An optical delay line was used to control the probe delay time, and a mechanical chopper was used to modulate the pump pulse, providing alternating probe-only and pump-probe pulses, allowing us to measure pump-induced absorption changes. Measurements at different molecular concentrations were performed, adjusting the pump fluence in order to maintain an approximately constant photon density (i.e. pump photons per LFO molecule) $r = kN_\gamma/N$, where $N$ is the total number of molecules in the excitation volume, $N_\gamma$ is the total number of pump laser photons, and $k$ is the fraction of them that actually reach the active layer of the microcavity. We estimate from the reflectivity data that only 6% to 8% of the initial pump excitation enters the cavity. We conducted our experiment in air at room temperature.

**Results**

We first show that ultrafast transient-absorption spectroscopy can monitor the population of excited molecules, even in a cavity, by comparing the control film and the microcavity spectra as shown in Fig. 2a. A representative control film $\Delta T/T$ spectrum is shown for a probe delay time of 1.0 ps, and the $\Delta R/R$ spectra of the microcavities are shown at a delay of 1.25 ps. (Further data is given in the supplementary information). We found the control film spectra at all concentrations to show two positive bands around 530 and 577 nm, which both reflect excited state populations. By comparison with the spectra in Fig.1b, we attribute the 530 nm band to ground state bleaching (GSB) – i.e. suppression of absorption due to molecules already being in their excited state. The 577 nm band instead corresponds to stimulated emission (SE) by excited molecules. For each of the microcavity spectra, we have a single prominent peak, which corresponds to the transient signal filtered by the cavity mode. This implies that the time-dependent transient reflectivity signal is proportional to the change in the number of excited molecules created by the pump(*32*), i.e. $\frac{\Delta R}{R}(t) \propto N_\uparrow(t)$. Since the energy stored in the molecules is also proportional to the number of excited molecules $E(t) \propto N_\uparrow(t)$, we can thus monitor the stored energy. While experiment directly provides the time dependence, estimating the absolute scale of energy density requires multiplying $\Delta R/R$ by a time-independent constant. Estimating this constant from first principles is challenging, so we instead extract it through fitting to the theoretical model which is discussed below. This fitting is discussed in section S3 of the Supplementary Material. We also note that two of the microcavity spectra show a negative $\Delta R/R$ band, which results from the change in the refractive index induced by the pump pulse (*33*).

Figure 2b shows the experimental values for the time-dependent stored energy density. In all microcavities studied, the energy density undergoes a rapid rise followed by slow decay. The timescale of the rapid rise varies with concentration. We adjust the laser power to fix photon density $r$ across comparable microcavities, and compare behaviour with different LFO concentrations. Details of how $r$ is estimated are provided in the supplementary information. We found that to achieve a sufficiently high signal-to-noise ratio, it was not possible to compare all microcavities at the same $r$ value; instead, a constant $r$ value was maintained for matched structures. Specifically, measurements were made on microcavities with LFO concentrations of 10%, 5% and 1% with approximately constant $r \simeq 0.14$ (respectively labelled A1, A2, and A3), and 1% and 0.5% with $r \simeq 2.4$ (labelled B1 and B2).

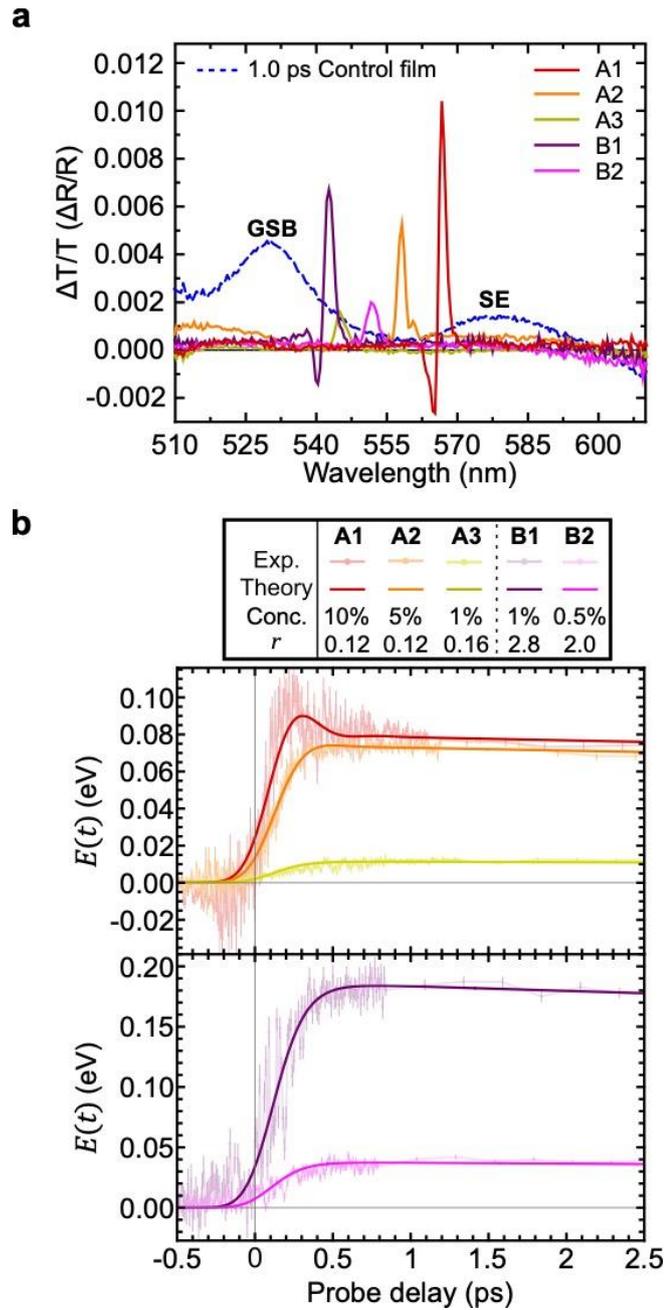

**Fig. 2 Experimental demonstration of superextensive charging. (a)** Differential-transmittivity ($\Delta T/T$) spectra for the control film (at 1% LFO concentration) at a probe delay time of 1.0 ps, and the differential-reflectivity ($\Delta R/R$) spectra for the microcavities at 1.25 ps probe delay. **(b)** Temporally resolved energy density of the microcavities shows that rise time decreases as stored energy density increases, indicating superextensive charging. A1, A2 and A3 label results for microcavities containing LFO at concentrations of 10%, 5% and 1%, as the ratio of pump photons to molecules is kept approximately constant at $r \simeq 0.14$. B1 and B2 label

measurements for LFO at concentrations of 1% and 0.5%, with $r \simeq 2.4$. The use of two different $r$ values was necessary to achieve a sufficiently high-signal-to-noise ratio. Points mark the experimental data, while continuous solid lines are the results of the theoretical model, with parameters given by a chi-squared minimisation of the experimental data. Experimental uncertainties are estimated from the point-to-point variance of the data.

Overlaying the experimental data are the corresponding theoretical predictions (see Theoretical Model section). To account for both the instrument response time (~20 fs) and the cavity photon lifetime (which was estimated as discussed in the supplementary information), the theoretical curves are convolved with a Gaussian response function with a full-width half maximum of ~120 fs. There is good agreement between the experimental data and the corresponding theoretical curves.

To obtain energetic dynamics, we take away the response function from the theoretical fit, as shown in Supplementary Figure S15. Table 1 summarises the rise time or the time to reach half maximum energy ($\tau$), the peak stored energy density ($E_{\max}$), and the charging rate or peak charging power density [$P_{\max} = \max(dE / dt)$]. These are extracted from the theoretical fit to the data presented in Figs. 2b. We see that $\tau$ decreases with $N$, whilst $E_{\max}$ and $P_{\max}$ increase with $N$. Recalling that $E_{\max}$ and $P_{\max}$ are the stored energy and charging power *per molecule*, this indicates superextensive behaviour. The scaling with $N$ is not the same across all experiments and in Supplementary Table S2 we summarise the different scaling.

Our results demonstrate that as the number of molecules in the microcavity increases, its charging power density remarkably increases. This means that it takes less time to charge a single microcavity containing $N$ molecules, than it would to charge $N$ single-molecule microcavities, even if the latter were charged simultaneously.

Furthermore, one microcavity with $N$ molecules would store more energy than $N$ microcavities, where each contained a single molecule. These superextensive properties are the key experimental findings of our work, and are supported by the theoretical modelling presented in the next section.

| Exp. | $N(\times 10^{10})$ | $\tau$ (ps) | $E_{max}$ (eV) | $P_{max}$ (eV / ps) |
|---|---|---|---|---|
| *A1* | 16.0 | 0.094 | 0.108 | 0.791 |
| *A2* | 8.1 | 0.120 | 0.076 | 0.412 |
| *A3* | 1.6 | 0.118 | 0.011 | 0.060 |
| *B1* | 0.16 | 0.114 | 0.184 | 1.008 |
| *B2* | 0.21 | 0.105 | 0.037 | 0.221 |

**Table 1: Summary of the experimental results.** In experimental groupings A1, A2, A3 and B1, B2, the number of molecules ($N$) increase whilst the ratio of photons to molecules remains constant ($r \approx 0.105$ and 2.4 respectively). The rise time $\tau$, is defined by the time to reach $E_{max}/2$, where $E_{max}$ is the peak stored energy per molecule or energy density. The charging rate $P_{max} = \text{Max}(dE/dt)$, is the peak charging power per molecule or charging power density.

**Theoretical Model**

The experimental dynamics can be reproduced by modelling, with the Lindblad master equation, the $N$ TLSs in an optical cavity with light-matter coupling strength $g$, a driving laser with a Gaussian pulse envelope and peak amplitude $\eta_0$, and three decay channels corresponding to the cavity decay ($\kappa$), TLS dephasing ($\gamma^z$), and TLS relaxation ($\gamma^-$). To solve this many-body Lindblad master equation, we make use of the cumulant expansion(*34-36*), with model parameters given by a chi-squared minimisation of the experimental data. Experimental uncertainties are estimated from the point-to-point variance of the data. Further details can be found in the Methods and Supplementary Information.

From our cumulant expansion simulations, we show how $\tau, E_{max}$, and $P_{max}$ vary as a function of $N$ in Fig. 3a and b. The interplay amongst the decay channels, driving laser, and cavity couplings give rise to a rich set of behaviour. We identify three regimes: decay-dominated at small $N$, and coupling-dominated at large $N$, along with a crossover regime between them. The system exhibits superextensive energy density scaling in the decay-dominated regime, and subextensive charging time in the coupling-dominated regime. In the crossover regime, the system exhibits both superextensive energy density scaling and subextensive charging times. Importantly, charging power density is superextensive in all regimes.

Figures 3c and d show the typical time dependence in decay-dominated and coupling-dominated regimes, indicating how the model parameters affect the dynamics. In particular, the presence of the decay channels gives rise to ratchet states which are capable of absorbing but not emitting light(*37*), thereby allowing the energy to be stably stored. See Methods and Supplementary Information for further discussion on the operating regimes. Figure 3 is augmented with an animation of how the energetic dynamics changes with $N$ (see Supplementary Video).

Figures 3a and b provide an explanation for the different scaling of the observables with $N$ shown in Table 1. Specifically, A1 and A2 operate in the coupling-dominated regime, where $\tau$ scales slightly less than $N^{-1/2}$, $E_{max}$ scales slightly more than $N^0$, and $P_{max}$ scales slightly more than $N^{1/2}$. For the region between A2 and A3, the average scaling of $\tau$ falls between $N^0$ and $N^{-1/2}$, $E_{max}$ between $N^2$ and $N^0$, and $P_{max}$ between $N^2$ and $N^{1/2}$. As A2 is further in the coupling-dominated regime than A3 is in the decay-dominated regime, the average scaling values between A2 and A3 are skewed towards the coupling-dominated scalings. B1 and B2 operate in the cross-over regime, with an average scaling with $N$ that is between the decay-dominated and coupling-dominated scalings, as reflected in Table 1.

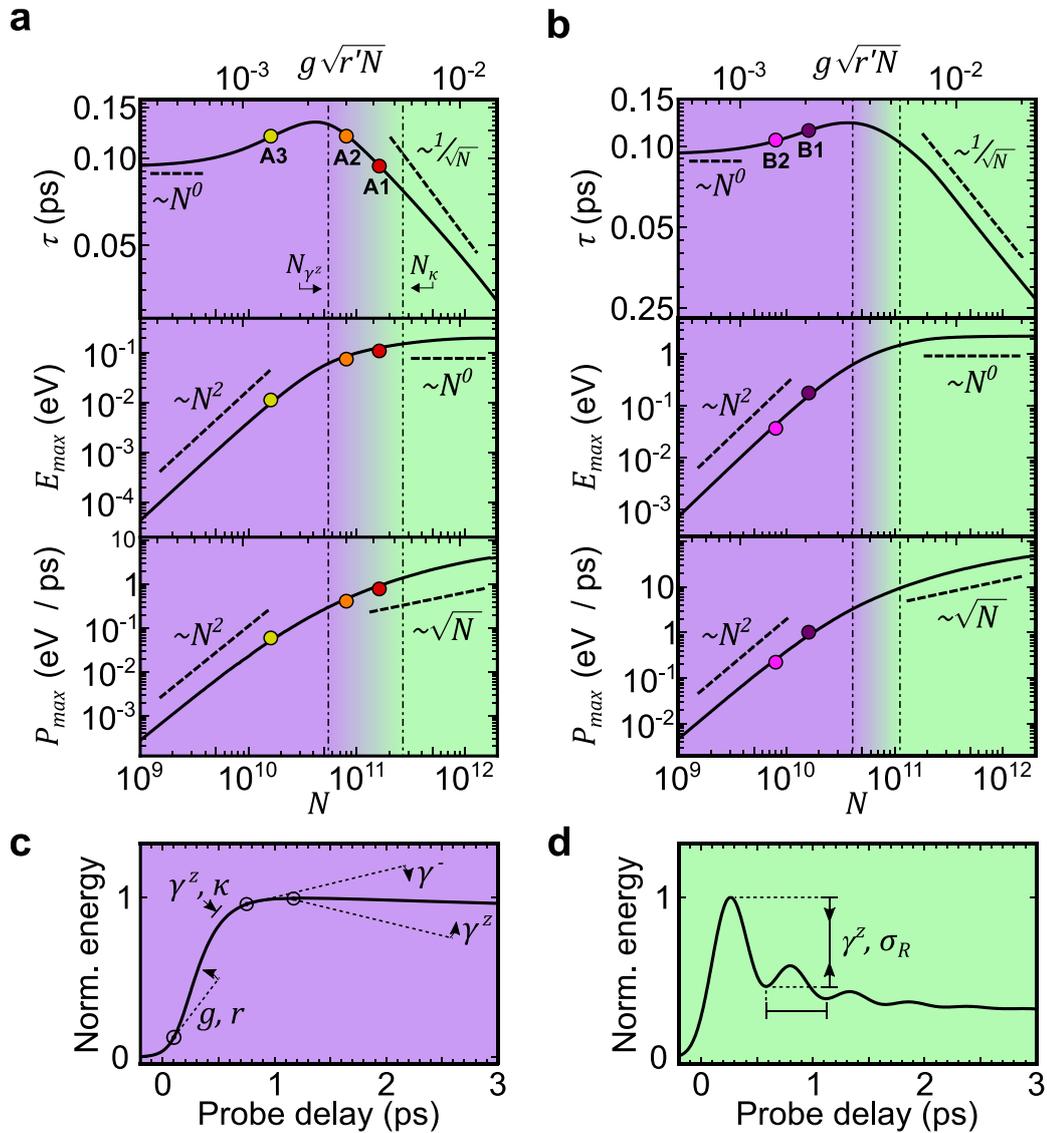

**Fig. 3: Charging dynamics as function of the number of molecules.** (**a**) and (**b**) show the theoretical model (solid line) for $r = 0.14$ and $2.4$, respectively. We show three operating regimes: decay-dominated (purple), coupling-dominated (green), and a decay-coupling-crossover regime. The decay-dominated regime is bounded by $N_\kappa < \kappa^2/g^2 r'$, and the coupling-dominated regime is bounded by $N_{\gamma^z} > \gamma^{z^2}/g^2 r'$, where $r' = \max(1, r)$. The coloured dots

indicate where the experiments sit on these curves. The uncertainty in $N$ is 10%, which is smaller than the dot size. (**c**) qualitatively depicts the effects of the model parameters in shaping the dynamics in the decay-dominated regime. (**d**) qualitatively depicts the effects of the additional model parameters in shaping the dynamics in the coupling-dominated regimes. $\sigma_R$ is the temporal width of the instrument response function.

## Summary

We have provided direct experimental evidence of superextensive energy storage capacity and charging in an organic microcavity by using ultrafast optical spectroscopy. Our realisation of a prototype Dicke QB highlights the fact that purely closed unitary dynamics is insufficient for realising a practical QB. The retention of energy requires finely-tuned decoherence processes, allowing the battery to charge quickly and yet discharge much more slowly. Such stabilisation of stored energy is a key step to exploit superextensive charging. Our observation of dephasing shows that realistic noisy environments can aid implementation and application of useful QBs. A challenge for future work is to explore further how concepts of ratchet states could keep a QB operating in the range of higher-lying energy states that are associated with maximum absorption enhancement, i.e. near the mid-point of the Dicke ladder(*37*).

We conclude by discussing the potential for future applications based on superextensive charging. One practical challenge noted above, is that quenching limits the performance of the QB at high concentrations. Overcoming this limitation requires careful choice of materials in order to suppress intermolecular quenching. We note that there are classes of materials where quenching is particularly suppressed. For example in proteins such as Green Fluorescent Protein (GFP)(*38*) the active chromophore is surrounded by a cage, which suppresses exciton-exciton quenching at high intensities.

Such materials might provide a route to allow the study of higher concentrations. Beyond energy storage, the key challenge for practical applications of this effect is its integration in devices where energy can be efficiently extracted and used. While our focus has been on the quantum advantage in charging, there do exist approaches to efficiently extract energy. For example, this may be achieved by including charge transport layers between the active layer and the cavity layers(*39*). The transport layers allow charge separation of the excitons as well as preventing recombination. This transforms the top cavity layer into a cathode and the bottom cavity into an anode, giving rise to an electric current. As such, our work provides a direct path for the integration of the superextensive energy absorption process in an organic photovoltaic device. The fast dynamics of such a device may also be useful as an optical sensor in low-light conditions, or potentially for energy harvesting applications(*40-43*). More generally, the idea of superextensive charging may have wide-reaching consequences for sensing and energy capture and storage technologies.

**Materials and Methods**

**Device fabrication**

The microcavities constructed consist of a thin layer of LFO (Kremer Pigmente) dispersed in a PS (Sigma-Aldrich, average molecular weight ~ 192,000) matrix. The bottom DBR consisted of 10 pairs of $SiO_2$/$Nb_2O_5$ and were fabricated using a mixture of thermal evaporation and ion-assisted electron beam deposition by Helia Photonics Ltd. Solutions of LFO dissolved in 25 mg/mL PS in dichloromethane were prepared at 0.5%, 1%, 5%, and 10% concentration by mass. Each LFO solution was then spin-coated on top of the bottom DBR to produce a thin film with an approximate thickness of 185 nm. An 8-pair DBR was then deposited on top of the LFO layer using electron beam deposition. With this pair of mirrors, the reflectivity was >99% in the spectral region of interest(*44*).

The diluted molecules are expected to be isolated at low concentration 0.1 – 1%, but at higher dye concentrations, the 0-0 emission transition red-shifts by a few nm and the second peak increases in intensity due to aggregation of the dye molecules. This is evident in Supplemental Fig. S2a and b, with additional broader features observed at longer wavelengths, which we assign to intermolecular states such as excimers.

The 0.5% and 1% cavities lie in the weak-coupling regime, i.e. no polaritonic splitting could be seen in the cavity reflectivity spectrum, as shown in Supplemental Fig. S2. For the 5% cavity, we see a weak anti-crossing feature in the reflectivity spectrum (a small kink near the crossing), indicating operation in the intermediate coupling regime. The 10% cavity operated in the strong-coupling regime, showing a Rabi splitting of around 100 meV around the 0-0 transition (along with intermediate-coupling between the cavity mode and the 0-1 transition).

Supplementary Fig. S3 shows a transfer matrix simulation of the electric field distribution of the 1% cavity (the cavities exhibit similar distributions).

**Pump-probe spectroscopy**

Probe and pump pulses were generated by a broadband non-collinear optical parametric amplifier (NOPA). The NOPA was pumped by a fraction (450 µJ) of the laser beam generated by a regeneratively amplified Ti:Sapphire laser (Coherent Libra) producing 100 fs pulses at 800 nm at a repetition rate of 1 kHz. A pair of chirped mirrors were placed at the output of the NOPA to compensate for temporal dispersion, and by using 7 'bounces' we were able to generate pulses with a temporal width below 20 fs. The laser beam was then split by a beam-splitter, with the probe being delayed via a translation stage and the pump being modulated mechanically using a chopper at 500 Hz.

**Lindblad master equation**

As noted above, we find the experimental behaviour is well reproduced by the dynamics of the Dicke model, a model of a microcavity photon mode coupled to two-level systems representing the molecules. As further discussed in the supplement, such a model is generally an approximation for organic molecules, but for some systems can become a very accurate approximation in the limit of low temperatures(*45*).

The open driven nature of the experimental system is modelled with the Lindblad master equation,

$$\dot{\rho}(t) = -\frac{i}{\hbar}[H(t), \rho(t)] + \sum_{j=1}^{N}(\gamma^z \mathcal{L}[\sigma_j^z] + \gamma^- \mathcal{L}[\sigma_j^-]) + \kappa \mathcal{L}[a] , \quad (2)$$

where $\rho(t)$ is the density matrix and $\mathcal{L}[O] \equiv O\rho\,O^\dagger - \frac{1}{2}O^\dagger O\rho - \frac{1}{2}\rho\,O^\dagger O$ is the Lindbladian superoperator. $a^\dagger$ and $a$ are the cavity photon creation and annihilation operators, and $\sigma_j^{x,y,z}$ are the Pauli spin matrices for each molecule, with the raising and lowering spin operators defined as $\sigma_j^\pm = (\sigma_j^x \pm i\sigma_j^y)/2$ . There are three decay channels corresponding to the cavity decay ($\kappa$), dephasing ($\gamma^z$), and relaxation rate ($\gamma^-$) of the individual TLSs. The Hamiltonian for the LFO molecules in cavity is modelled as a collection of non-interacting TLSs with characteristic frequency $\omega$ equal to that of the cavity mode, and resonantly coupled to the cavity with strength $g$. The molecules are driven by a laser described by a Gaussian pulse envelope $\eta(t) = \frac{\eta_0}{\sigma\sqrt{2\pi}}e^{-\frac{1}{2}\left(\frac{t-t_0}{\sigma}\right)^2}$, and a carrier frequency $\omega_L$. We work in the frame of the laser carrier frequency, and so write

$$H(t) = \hbar\Delta a^\dagger a + \sum_{j=1}^{N}\left[\frac{\hbar\Delta}{2}\sigma_j^z + g(a^\dagger\sigma_j^- + a\sigma_j^+)\right] + i\hbar\eta(t)(a^\dagger - a) , \quad (3)$$

where $\Delta = \omega - \omega_L$ is the detuning of the cavity frequency from the laser driving frequency. The LFO molecules are initially in the ground state, and the laser is on-resonance ($\Delta = 0$).

**Cumulant expansion**

The energy density of the cavity containing identical molecules with transition energy $\omega$ is $E(t) = \frac{\hbar\omega}{2}[\langle\sigma^z(t)\rangle + 1]$. In general, the equation of motion $(\partial/\partial t)\langle\sigma^z\rangle = Tr[\sigma^z\dot{\rho}]$ depends on both the first order moments $\langle\sigma^{x,y,z}\rangle$ and $\langle a\rangle$ as well as higher order moments, leading to a hierarchy of coupled equations. Within mean field theory, the second order moments are factorised as $\langle AB\rangle = \langle A\rangle\langle B\rangle$ which closes the set of equations at first order. This approximation is valid at large $N$, as corrections scale as $1/N$. To capture the leading order effects of finite-sizes we make a second-order cumulant expansion(*34-36*), i.e. we keep second-order cumulants $\langle\langle AB\rangle\rangle = \langle AB\rangle - \langle A\rangle\langle B\rangle$ and assume that the third-order cumulants vanish, which allows us to rewrite third-order moments into products of first and second-order moments(*46*). In our experiments, the number of molecules in the cavity is large ($>10^{10}$) and we find higher order correlations are negligible. We give the equations of motion up to second order in the Supplementary Information.

**Operating regimes**

The decay-dominated (purple region in Fig. 3a and b) regime occurs when the collective light-matter coupling is weaker than the decay channels, $g\sqrt{Nr'} < \{\kappa, \gamma^z, \gamma^-\}$, where $r' = \max(1, r)$. In this regime, the time scale of cavity dynamics is slow relative to the decay rate. Fig. 3c shows a typical time dependence in this regime, indicating how the model parameters affect the dynamics. In this regime, the increase in the effective coupling relative to the decay strength sees an $N^2$ superextensive scaling of the energy and power density, while rise time remains constant. Experiment A3 operates near the boundary of this regime (Fig. 3a).

In the coupling-dominated (green region in Fig. 3a and b) regime, the effective collective light-matter coupling $g\sqrt{Nr'} > \{\gamma^z, \gamma^-, \kappa\}$, dominates over the decay channels. In this regime, the time scale of cavity dynamics is fast relative to the decay rate, and we observe $\sqrt{N}$-superextensive power scaling and $1/\sqrt{N}$ dependence of rise time, while the maximum energy density remains constant. While power scaling is superextensive in both regimes, the origin of this differs: for the decay-dominated regime this is the result of the superextensive energy scaling, while for the coupling-dominant regime it is the result of a superextensive decrease in the rise time. Experiments A1 and A2 operates in this regime (Fig. 3a).

In the crossover between the regimes (purple-green), the collective coupling falls between the cavity decay rate and the TLS dephasing rate, $\{\kappa, \gamma^-\} < g\sqrt{Nr'} < \gamma^z$. In Fig. 3a and b, $\gamma^-$ is small such that $g\sqrt{Nr'} \gg \gamma^-$ for all values of $N$, and so there is no boundary labelled for this decay rate. In this case, capacity and rise-time can simultaneously scale super- and subextensively, but at a rate slower than in the decay and coupling-dominated regimes, respectively. Experiments B1 and B2 operate in this regime (Fig. 3b).

**Decay and coupling rates**

The parameters needed in the theory calculations are the cavity leakage rate $\kappa$, the dephasing rate $\gamma^z$, the non-radiative decay rate $\gamma^-$, the interaction strength $g$, and the temporal width of the instrument response function, $\sigma_R$. Note that the temporal width of the pump pulse is fixed at $\sigma = 20$ fs. For the dephasing rate, we note that as one enters the strong-coupling regime, exciton delocalisation suppresses the effect of dephasing($47$). To approximately capture this effect, we assume that the dephasing rate scales with the number of molecules as $\gamma^z = \gamma_0^z \left( {N_{5\%}}/{N} \right)$ where $\gamma_0^z$ is taken to be constant and $N_{5\%}$ is the number of molecules in the 5% cavity. The experimental uncertainty in $N$ is estimated to be 10%. The cavity lifetime $T$ comes into the model in both $\sigma_R = T$ and the cavity leakage rate $\kappa = 1/T$. From transfer matrix modelling on the 1% and 0.5% cavities (where polariton effects are small) we estimate that $T \approx 306$ fs. However, based on the measured finesse of the cavities we estimate that $T = 120$ fs. Transfer matrix modelling assumes perfectly smooth mirrors, whilst measured finesse includes inhomogeneous broadening effects, neither of which we want to include in $\kappa$ and $\sigma_R$. In the following optimisation, we therefore assume that $T \in [120, 306]$ fs, with lower values more likely due to transfer matrix calculations being prone to error.

For $T$ values within this range, the remaining three parameters in the model ($\gamma_0^z, \gamma^-$ and $g$) were found through a global chi-squared optimisation, simultaneously optimising over all experiments. Uncertainties in these fitting parameters were then estimated by using the reduced $\tilde{\chi}^2$ distribution to find the 68% confidence interval of the model parameters. This corresponds to the range $\tilde{\chi}^2 \leq \tilde{\chi}^2_{min} + \Delta$, where for a three parameter optimisation and $k$ total data points $\Delta \approx 3.51/(k-3)$.($48$) In the Supplementary Information we present a figure showing minimum reduced chi-squared value as a function of $T$, and for each point we show the optimal set of parameters ($\gamma_0^z, \gamma^-$, $g$) along with the 68% confidence intervals. From this, and by comparison of the experimentally measured and theoretically calculated reflectivity for each parameter set, we concluded that the lifetime most representative of the data was $T = 120$ fs with $\gamma^- =$

$(0.0141^{+0.0031}_{-0.0024})$ meV, $g = (10.6^{+2.2}_{-1.3})$ neV and $\gamma_0^z = (1.68^{+0.25}_{-0.18})$ meV. See Supplementary Information for more details.

**Acknowledgements**

We thank the U.K. EPSRC for part funding this research via the Programme Grant 'Hybrid Polaritonics' (EP/M025330/1). We also thank the Royal Society for an International Exchange Grant (IES\R3\170324) 'Development of BODIPY dyes for strongly coupled microcavities'. K.M. thanks the University of Sheffield for a PhD studentship via the EPSRC DTG account x/012169-15. D. R. acknowledges studentship funding from EPSRC under grant no. EP/L015110/1. T.V. and L.G. thank the Regione Lombardia Funding project IZEB. J.Q.Q. acknowledges the Ramsay fellowship and the Centre for Nanoscale BioPhotonics Family Friendly Fund, for financial support of this work. We thank Caspar Clark and Ross Preston at Helia Photonics Ltd for fabricating the bottom DBRs. We also thank Richard Grant for the measurement of concentration-dependent photoluminescence quantum-yield of the LFO. **Author contributions:**

J.Q.Q. conceived and managed the project. K.E.M and D.G.L. contributed to the fabrication of the Dicke QBs. L.G, K.E.M., G.C. and T.V. contributed to the measurement of the Dicke QBs. D.M.R, J.Q.Q., B.W.L, E.M.G. and J.K. contributed to the theoretical analysis. All authors contributed to discussion of the results and the writing of the manuscript. **Data and material availability:** The research data supporting this publication can be accessed at https://doi.org/10.17630/66875381-317e-4d6c-b884-d069547301ea. **Competing Interests:** All authors declare that they have no competing interests.

# Supplementary Material
# Superabsorption in an organic microcavity: Towards a quantum battery

James Q. Quach,[1, *] Kirsty E. McGhee,[2] Lucia Ganzer,[3] Dominic M. Rouse,[4] Brendon W. Lovett,[4] Erik M. Gauger,[5] Jonathan Keeling,[4] Giulio Cerullo,[3] David G. Lidzey,[2] and Tersilla Virgili[3, †]

[1]*Institute for Photonics and Advanced Sensing and School of Chemistry and Physics, The University of Adelaide, South Australia 5005, Australia*
[2]*Department of Physics and Astronomy, University of Sheffield, Hicks Building, Hounsfield Road, Sheffield S3 7RH, U.K.*
[3]*Istituto di Fotonica e Nanotecnologia – CNR, IFN - Piazza Leonardo da Vinci 32, 20133 Milano, Italy*
[4]*SUPA, School of Physics and Astronomy, University of St Andrews, St Andrews KY16 9SS, United Kingdom*
[5]*SUPA, Institute of Photonics and Quantum Sciences, Heriot-Watt University, Edinburgh EH14 4AS , United Kingdom*

## S1. CHARACTERISATION OF SAMPLES AND CALIBRATION MEASUREMENTS

This section presents further details of the properties of the fabricated microcavities, and measurements used to calibrate the results in the main text.

*a. Quenching at large concentrations* Figure S1 shows the photoluminescence quantum yield (PLQY) as a function of dye concentration. At large concentration the yield drops to zero; this provides an upper limit on the concentration that can be studied in experiment. The photoluminescence measurements were taken using a Coherent Mira 900 laser operating at 400 nm with a repetition rate of 80 MHz. The laser beam was focused onto the surface of a sample placed at the centre of an integrating sphere. The laser light and the emission from the sample were scattered by the diffuse interior of the sphere and collected by an optic fibre, which was coupled to an Andor Shamrock SR-303i-A CCD spectrometer. Spectra were taken for different concentration LFO films, as well as a blank glass substrate in order to calculate the proportion of laser light absorbed by the LFO samples.

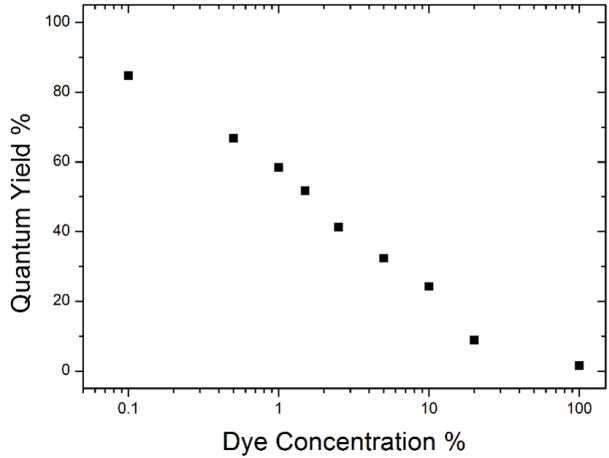

FIG. S1. Photoluminescence quantum yield as a function of LFO concentration.

---

* quach.james@gmail.com
† tersilla.virgili@polimi.it

*b. Film and microcavity spectra*   Figure S2 shows the molecular absorption and emission spectra, and examples of the microcavity reflectivity spectra. Panels (a,b) show the properties of the bare molecular film. These show the small Stokes shift between absorption and photoluminescence, and also show how high film concentrations modify the photoluminescence spectrum, consistent with the reduced PLQY shown above. The microcavity reflectivity spectra (c,d) show the crossover from weak- to strong-coupling as the concentration is increased.

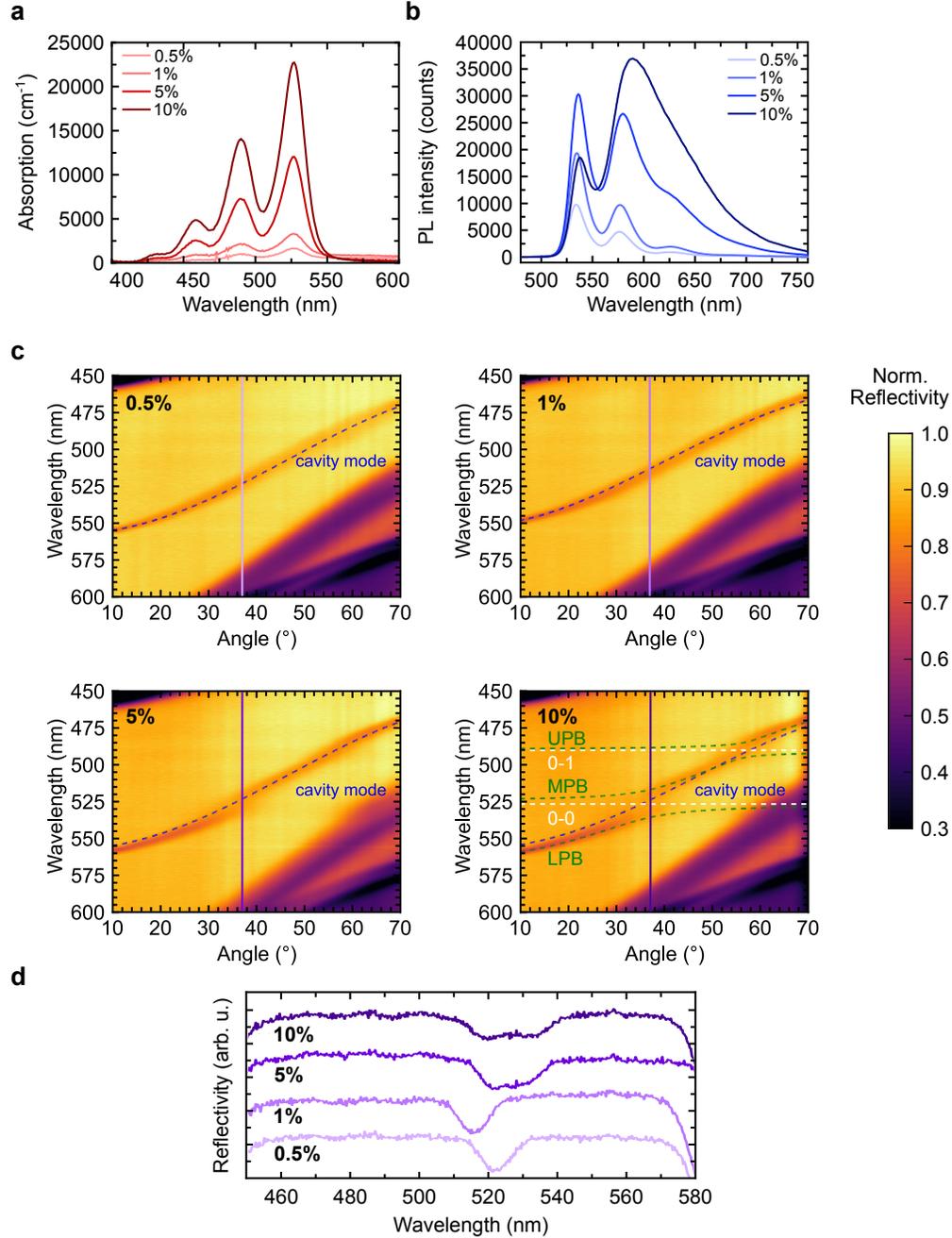

FIG. S2. **Absorption, photoluminescence of the LFO films and reflectivity spectra of the microcavities.**
(a) Absorption and (b) photoluminescence spectra for the 0.5%, 1%, 5%, and 10% LFO-concentration films. (c) Reflectivity spectra for 0.5%, 1%, 5%, and 10% LFO-concentration microcavities. UPB, MPB, and LPB label the upper, middle, and lower polariton branches, respectively. Also indicated are the 0-0 and 0-1 transition wavelengths. (d) is a slice of the reflectivity spectra at 37°. The single dip in the 0.5% and 1% concentration spectra indicate the weak-coupling regime. The double dip seen in the 10% concentration spectra, represent the polaritonic states, indicating the strong-coupling regime. The 5% concentration spectrum represents a situation intermediate between a single and double dip, indicating an intermediate-coupling regime.

*c. Transfer matrix calculations* To provide bounds on the cavity lifetime, separate from the measured cavity linewidth—which contains effects of inhomogeneous broadening—we make use of transfer matrix calculations of the cavity structure. These calculations, shown in Fig. S3, also enable one to visualise the electric field intensity in the microcavity structure. These calculations give a designed cavity lifetime of 306fs, which serves as an upper bound of the actual cavity lifetime.

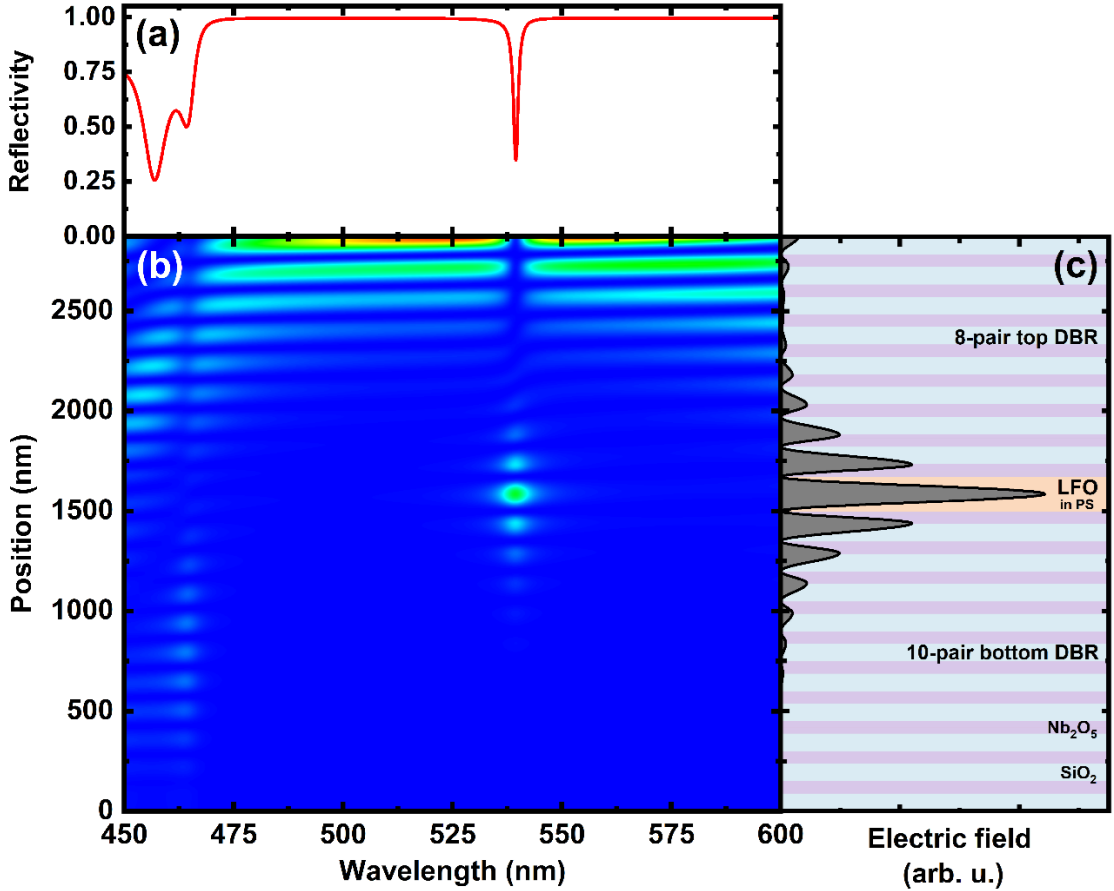

FIG. S3. **Transfer matrix simulation of the electric field distribution for the 1% cavity.** (a) shows the cavity reflectivity, (b) the spectrally-resolved electric field amplitude, and (c) the electric field amplitude at the cavity mode wavelength. The shaded sections of (c) indicate the different materials which make up the cavity, with Nb2O5 in purple (refractive index, n = 2.25), SiO2 in blue (n = 1.52), and LFO in PS in orange (n = 1.60). All simulations were made using transfer matrix modelling at an angle of 20° to the cavity normal to maintain consistency with the transient reflectivity measurements.

*d. Pump-probe dynamics of bare films* For comparison to the pump-probe dynamics of the cavity shown in the main text, Figs. S4,S5 show the transient transmission spectra of the bare films. (Note that, as discussed in the main text, a transmission geometry is required for transient spectroscopy of the bare films).

Figure S4 compares the dynamics at 525 nm (ground state bleaching) and 571 nm (stimulated emission region) for the 1% concentration film. Unfortunately a coherent artefact, due to the degenerate pump-probe configuration, masks the time dynamics in the first 100 fs. However, the same rise and decay times are seen at both probe wavelengths, indicating that we are probing the same exciton population.

We also explored pump-fluence dependence of the bare films. No dependence on the excitation fluence was detected in any control film. This indicates the absence of bimolecular effects or multi-photon excitation. Figure S5 shows such data for all concentrations, at two different pump fluences. As in Fig. S4, a coherent artefact is present at zero delay due to the degenerate pump and probe spectra.

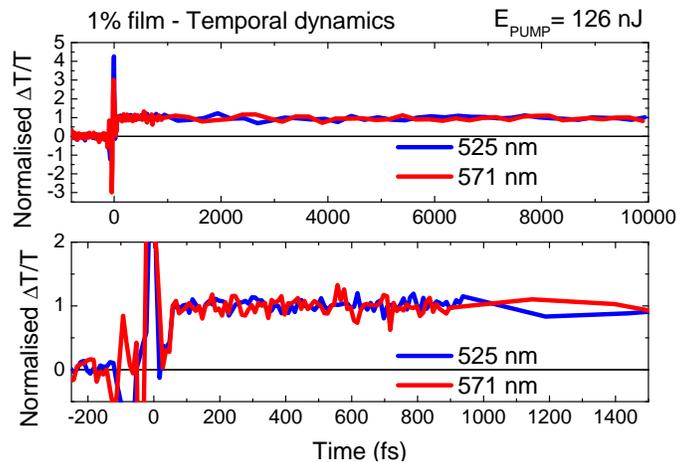

FIG. S4. **Dynamics of the control 1% control film at different wavelengths.** The two panels show two different time windows (top panel until 10 ps, bottom panel until 1.4 ps. The same behavior is observed for other film concentrations.

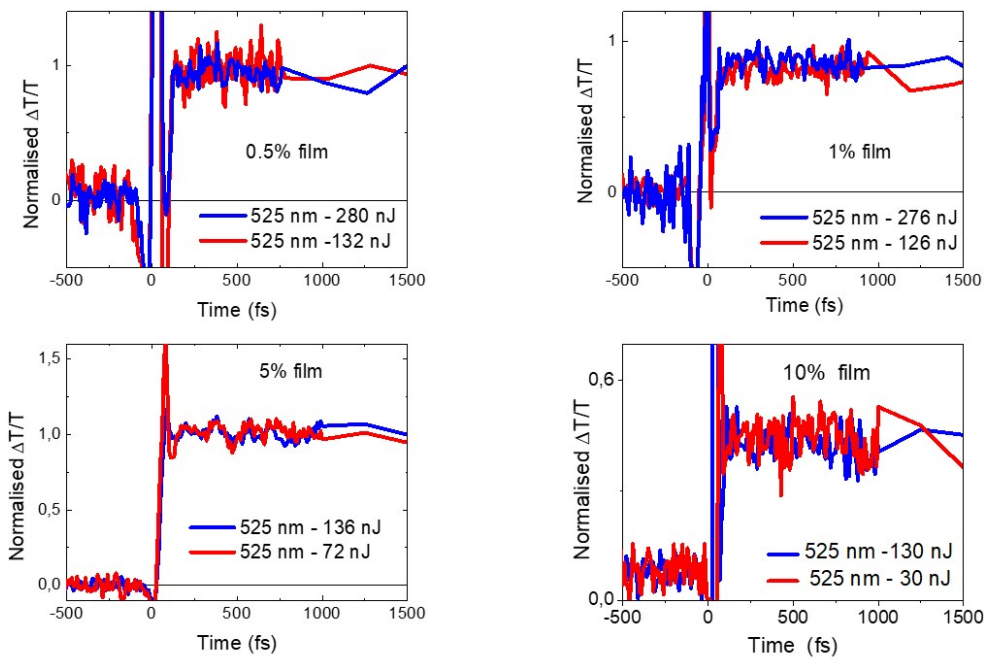

FIG. S5. **Dynamics of the control films, at different concentrations.** Each panel shows the dynamics for a different concentration, as indicated.

*e. Estimating number of molecules in film.* To determine the number of LFO molecules in the microcavities we study, we first determined the absorption cross-section of a single LFO molecule, $\sigma_{LFO}$. The transmission spectrum of a 0.1% solution of LFO in 25 mg/mL PS/dichloromethane in a 1 mm thick cuvette was measured using a Horiba Fluoromax 4 fluorometer with a xenon lamp. The absorption coefficient ($\alpha = n\sigma_{LFO}$) of the 0–0 transition was then calculated using the relation $T/T_0 = e^{-\alpha d}$, where $T/T_0$ is the fractional transmission of the xenon lamp at the 0–0 transition, $d$ is the cuvette thickness, $n$ is the number density of absorbing molecules in solution per unit volume, and $\sigma_{LFO}$ is the absorption cross-section of a single LFO molecule [49]. Using the known value of $n$ for this solution, $\sigma_{LFO}$ was calculated as $3.3 \times 10^{-16}$ cm$^2$. The transmission of the 10% LFO concentration in film was then measured to obtain $\alpha$ and hence $n$ (number density of molecules in the cavity active layer), using the measured value of $\sigma_{LFO}$, with $d$ (film thickness) measured using a Bruker DektakXT profilometer. This value was then multiplied by the area of the laser beam and $d$ to obtain $N$. Here we assume a uniform distribution in the active layer. $N$ for other concentrations were scaled accordingly.

*f. Estimating number of photons in cavity.* To estimate the number of photons entering the cavity in each different cavity, we consider the overlap between the pump spectrum and the cavity transmission. The number of photons in the cavity is given by multiplying the number of pump photons by $1 - R$, where $R$ is the reflectivity of the cavity: $n = N_\gamma(1 - R)$. An example of this is shown in Fig. S6, for the 1% cavity. Starting from the pump spectrum (yellow curve), by considering the reflectivity spectrum of the cavity (black line), we calculate the fraction of photons entering into the cavity (purple line). Table S1 shows the resulting estimates of photon numbers for each experiment.

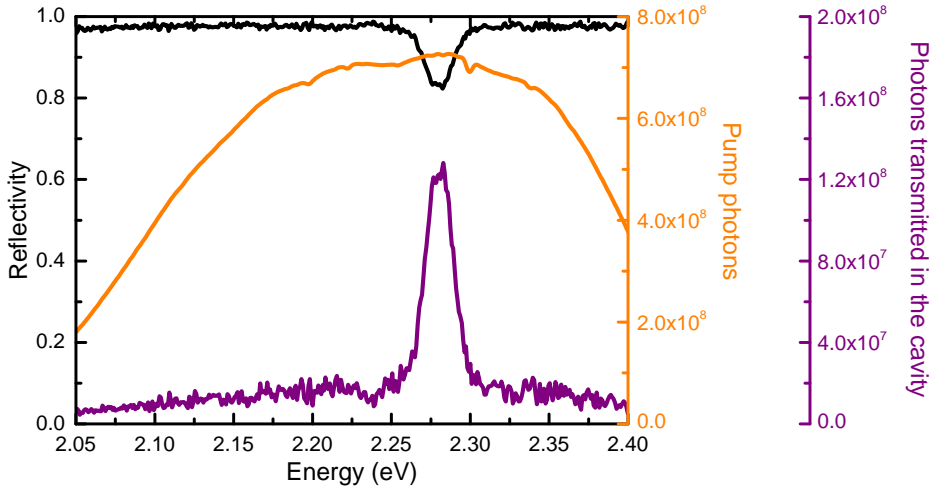

FIG. S6. **Calculating photon number**. Yellow curve (right axis): pump spectrum. Black curve (left axis) reflectivity spectrum of 1% cavity. Purple curve (right axis): photons transmitted into the cavity.

| Experiment | $N_{\mathrm{dye}}(\times 10^{10})$ | $N_{\mathrm{photon}}(\times 10^{10})$ |
|---|---|---|
| **A1** | 16.20 | 1.90 |
| **A2** | 8.08 | 0.98 |
| **A3** | 1.62 | 0.26 |
| **B1** | 1.62 | 4.53 |
| **B2** | 0.81 | 0.16 |

TABLE S1. Estimated photon number and molecule number for each experiment.

*g. Derivative features in the differential transmitivity* As seen in Fig. 2 of the main text, the transient signal shows both positive and negative features in the differential transmitivity. This can be understood from the existence of a derivative feature in the spectrum. Such a feature occurs if the pump causes an absorption peak to move in energy. In that case the change of absorption will see a decrease in absorption where the feature used to be, and an increase where the feature now is [33,50]. This leads to a feature that takes the form of the derivative of the original absorption peak with respect to energy, and which thus contains both negative and positive contributions. This structure is exactly what we observe, indicating that the feature corresponds to a resonance which moves with excited state population. This feature though coexists with other positive features in the transient reflectivity, arising from mechanisms such as ground state bleaching or stimulated emission. Whether a negative feature is seen depends on how these contributions compete and whether the feature is above the detection threshold.

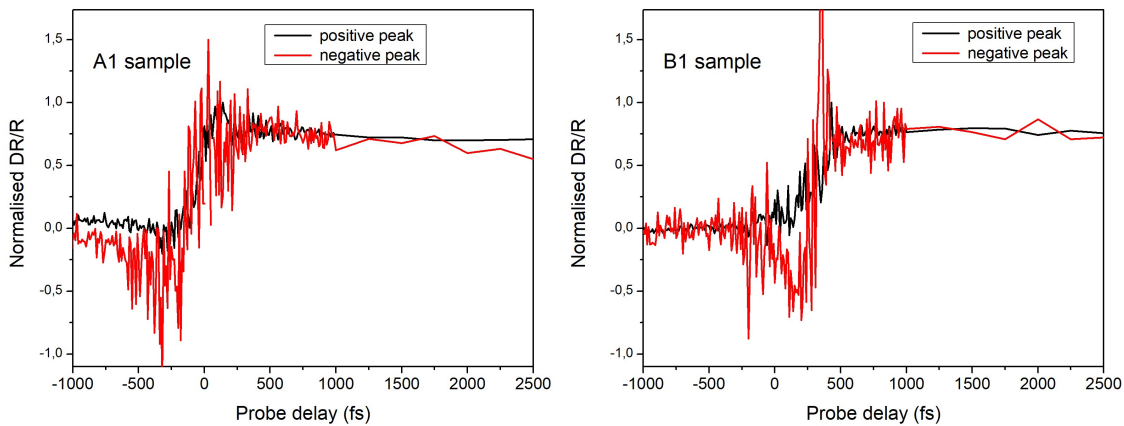

FIG. S7. Time evolution of positive and negative features in the differential transmitivity.

The presence of this derivative feature does not influence our determination of the energy stored in the molecules and the charging time. As shown in the Fig. S7, the time dependence of the negative and positive peaks is the same. (The negative feature does show more noise, as this signal is weaker compared to background noise.) Since both features show the same time evolution, they would both recover the same theoretical fits.

## S2. THEORETICAL MODELLING

In this section we describe our approach to modelling the system. We first provide a more detailed discussion of the model we use, and the physical origin of the terms involved. We then derive the equations of motion for the expectation values using a second-order cumulant approach.

### A. Details of theoretical model

As discussed in the main text, we model the experiments through a Dicke model—which describes $N$ two-level systems coupled to a photon mode. Describing organic molecules as two-level systems is an approximation, as it neglects both the role of rotational and vibrational molecular modes, as well as the existence of higher excited states of the molecules. It is known that this approximation can become valid in some limiting situations, such as at low temperatures [45]. While our experiments are performed at room temperature, we nonetheless expect the model to provide a reasonable approximation. This is because, as noted in the main text, the molecules we consider show a small Stokes shift, indicating vibrational dressing of optical transitions is weak. Moreover, as seen in the main text and discussed further below, our model matches the experimental results well without such additional features.

The Hamiltonian describing our system, including the external pump laser, takes the form (setting $\hbar = 1$):

$$H = \Delta_c a^\dagger a + \sum_{j=1}^{N} \left[ \frac{\Delta_a}{2} \sigma_j^z + g(a^\dagger \sigma_j^- + a\sigma_j^+) \right] + i\eta(t)(a^\dagger - a) \,, \tag{1}$$

where $a$ $(a^\dagger)$ is the photon annihilation (creation) operator, $\sigma_i^\alpha$ for $\alpha = x, y, z$ are the Pauli matrices. We write the Hamiltonian in the rotating frame of the pump laser, so $\Delta_a$ $(\Delta_c)$ is the energy detuning of the laser from the cavity (molecules), $g$ is the coupling strength of each molecule to the photon mode, $\eta(t)$ is the Gaussian profile of the pump laser. In the following we assume that the molecules and cavity are resonant $(\Delta_a = \Delta_c \equiv \Delta)$ and that $\Delta = 0$ unless specifically noted.

To account for dissipative processes, we consider the time evolution of the density matrix, including Lindblad terms for various dissipative processes:

$$\dot{\rho} = -i[H, \rho] + \kappa \mathcal{L}[a] + \sum_{j=1}^{N} \left( \gamma^z \mathcal{L}[\sigma_j^z] + \gamma^- \mathcal{L}[\sigma_j^-] \right) \,, \tag{2}$$

where $\mathcal{L}[X] = X\rho X^\dagger - (1/2)\{X^\dagger X, \rho\}$. The first term, with rate $\kappa$, describes loss of photons, due to the finite reflectivity of the cavity mirrors. The second term, with rate $\gamma^z$, describes dephasing of the molecular excitations. This term describes the effect of coupling between the molecular electronic state and vibrational degrees of freedom—vibrations both of the molecule and of the polystyrene matrix in which it is contained. The final term, $\gamma^-$, describes the decay of electronic excitation, due to emission into non-cavity modes, such that $1/\gamma^-$ would be the excited state lifetime in the absence of the cavity.

As noted in the main text, dephasing $\gamma^z$ plays a crucial role in the dynamics, with quite distinct behaviour occurring with and without dephasing. In particular, dephasing introduces transitions between the "bright" and "dark" molecular excited states. To understand these states, let us first consider states with a single excitation. The form of Eq. (1) shows that the cavity photon couples only to the totally symmetric molecular excited state, i.e. a state with equal weight and phase of excitation on all molecules. However, for $N$ molecules, $N$ excited states exist. The remaining $N - 1$ states are orthogonal to the coupling to light, and are known as "dark states". While we have introduced this for the space with a single excitation, a generalization to higher excited states exists. In this case, the language of superradiant and subradiant states is often used [1], with superradiant states referring to those that can be reached using the collective raising and lowering operators, $\sum_j \sigma_j^\pm$. Because the dephasing term acts on individual molecules, it describes a process that allows loss of phase coherence between different molecules. As such, this causes a transition from the optically bright state created by the laser, to an incoherent mixture of bright and dark states. Since the dark states do not couple to the cavity, this process is responsible for the asymmetry between collectively enhanced absorption, and the lack of collective enhancement of emission.

For $N$ identical molecules of energy $\omega_a$ (in the lab frame), the energy density stored on the molecules is given by

$$E(t) = \frac{\omega_a}{2} \left( \langle \sigma^z(t) \rangle + 1 \right) . \tag{3}$$

As such, in the following, our aim is to predict the time evolution of this quantity.

## B. Cumulant equations

To determine the time evolution of $\langle \sigma^z(t) \rangle = \mathrm{Tr}\left[ \sigma^z \rho(t) \right]$ we begin by writing down the first order expectation values of the system. We adopt the notation $C_a(t) \equiv \langle a(t) \rangle$ for photon operators, and $C_{\alpha=x,y,z}(t) \equiv \langle \sigma^\alpha(t) \rangle$ for spin operators, along with a similar notation for higher order expectations, e.g., $C_{ax}(t) \equiv \langle a\sigma^x(t) \rangle$. The equations of motion for the first order expectation values are

$$\partial_t C_a = -\left( i\Delta_c + \tfrac{1}{2}\kappa \right) C_a - \tfrac{1}{2}gN \left( iC_x + C_y \right) + \eta(t) , \tag{4}$$

$$\partial_t C_x = -\Delta_a C_y - 2g\mathrm{Im}\left[ C_{az} \right] - \gamma^{\mathrm{tot}} C_x , \tag{5}$$

$$\partial_t C_y = \Delta_a C_x - 2g\mathrm{Re}\left[ C_{az} \right] - \gamma^{\mathrm{tot}} C_y , \tag{6}$$

$$\partial_t C_z = 2g\left( \mathrm{Re}\left[ C_{ay} \right] + \mathrm{Im}\left[ C_{ax} \right] \right) - \gamma^- \left( C_z + 1 \right) , \tag{7}$$

where $\partial_t$ is short for $\frac{\partial}{\partial t}$, $\gamma^{\mathrm{tot}} = 2\gamma^z + \tfrac{1}{2}\gamma^-$, and for notational ease we have dropped the explicit time dependence of observables. As described in the main text, in mean field theory we would now set the second order cumulants to zero. These are defined as

$$\langle\langle AB \rangle\rangle = \langle AB \rangle - \langle A \rangle\langle B \rangle . \tag{8}$$

This would result in the usual decomposition of second order expectation values into products of first order ones, $C_{AB} = C_A C_B$, which is the assumption that molecule-molecule, molecule-photon, photon-photon and all higher order correlations are negligible. However, we instead derive equations of motion for the second order expectation values, capturing the leading order $1/N$ corrections to mean field theory. The second order photon correlations obey:

$$\partial_t C_{a^\dagger a} = -\kappa C_{a^\dagger a} - gN \left( i\mathrm{Im}\left[ C_{ax} \right] + \mathrm{Re}\left[ C_{ay} \right] \right) + 2\eta(t)\mathrm{Re}\left[ C_a \right] , \tag{9}$$

$$\partial_t C_{aa} = -\left( 2i\Delta_c + \kappa \right) C_{aa} - gN \left( iC_{ax} + C_{ay} \right) + 2\eta(t)C_a , \tag{10}$$

while molecule-photon correlations follow:

$$\partial_t C_{ax} = -\left( i\Delta_c + \tfrac{1}{2}\kappa + \gamma^{\mathrm{tot}} \right) C_{ax} - \Delta_a C_{ay} - i\frac{g}{2}\left[ 1 + (N-1) \right] C_{xx}$$
$$- \frac{g}{2}\left[ iC_z + (N-1) C_{xy} \right] + ig\left( C_{aaz} - C_{a^\dagger az} \right) + \eta(t)C_x , \tag{11}$$

$$\partial_t C_{ay} = -\left( i\Delta_c + \tfrac{1}{2}\kappa + \gamma^{\mathrm{tot}} \right) C_{ay} + \Delta_a C_{ax} - i\frac{g}{2}\left[ -iC_z + (N-1) C_{xy} \right]$$
$$- \frac{g}{2}\left[ 1 + (N-1) C_{yy} \right] - g\left( C_{aaz} + C_{a^\dagger az} \right) + \eta(t)C_y , \tag{12}$$

$$\partial_t C_{az} = -\left( i\Delta_c + \tfrac{1}{2}\kappa \right) C_{az} - \gamma^- \left( C_{az} + C_a \right) - \frac{g}{2}\left[ -iC_x + (N-1) C_{yz} \right]$$
$$- i\frac{g}{2}\left[ iC_y + (N-1) C_{xz} \right] + g\left( C_{aay} + C_{a^\dagger ay} \right) - ig\left( C_{aax} - C_{a^\dagger ax} \right) + \eta(t)C_z . \tag{13}$$

These now depend on third order expectation values, some of which contain multiple Pauli operators. We must note that these terms indicate Pauli operators representing different molecules and so commute — we have already taken into account the cases where the Pauli operators correspond to the same molecule by using the Pauli algebra $\sigma^\alpha\sigma^\beta = \mathbb{1}\delta^{\alpha\beta} + i\sigma^\gamma\epsilon^{\alpha\beta\gamma}$. The molecule-molecule expectation values for the same Pauli operator acting on different molecules are

$$\partial_t C_{xx} = -2\Delta_a C_{xy} - 4g\mathrm{Im}\left[ C_{axx} \right] - 2\gamma^{\mathrm{tot}} C_{xx} , \tag{14}$$

$$\partial_t C_{yy} = 2\Delta_a C_{xy} - 4g\mathrm{Re}\left[ C_{ayz} \right] - 2\gamma^{\mathrm{tot}} C_{yy} , \tag{15}$$

$$\partial_t C_{zz} = 4g\left( \mathrm{Im}\left[ C_{axz} \right] + \mathrm{Re}\left[ C_{ayz} \right] \right) - 2\gamma^- \left( C_{zz} + C_z \right) . \tag{16}$$

Finally, the molecule-molecule expectation values for different Pauli operators acting on different molecules are

$$\partial_t C_{xy} = \Delta_a \left( C_{xx} - C_{yy} \right) - 2g \left( \mathrm{Re} \left[ C_{axz} \right] + \mathrm{Im} \left[ C_{ayz} \right] \right) - 2\gamma^{\mathrm{tot}} C_{xy} \,, \tag{17}$$

$$\partial_t C_{xz} = -\Delta_a C_{yz} + 2g \left( \mathrm{Re} \left[ C_{axy} \right] + \mathrm{Im} \left[ C_{axx} \right] - \mathrm{Im} \left[ C_{azz} \right] \right) - \gamma^{\mathrm{tot}} C_{xz} - \gamma^- \left( C_{xz} + C_x \right) \,, \tag{18}$$

$$\partial_t C_{yz} = \Delta_a C_{xz} + 2g \left( \mathrm{Re} \left[ C_{ayy} \right] - \mathrm{Re} \left[ C_{azz} \right] + \mathrm{Im} \left[ C_{axy} \right] \right) - \gamma^{\mathrm{tot}} C_{yz} - \gamma^- \left( C_{yz} + C_y \right) \,. \tag{19}$$

In principle one can continue to write equations of motion for increasingly higher orders of expectation values, however, at large $N$, most essential physics is obtained at second order. We therefore truncate the cumulant expansion by setting third order cumulants to zero. These are defined as

$$\langle\langle ABC \rangle\rangle = \langle ABC \rangle - \langle AB \rangle \langle C \rangle - \langle A \rangle \langle BC \rangle - \langle AC \rangle \langle B \rangle + 2 \langle A \rangle \langle B \rangle \langle C \rangle \,, \tag{20}$$

and so setting $\langle\langle ABC \rangle\rangle = 0$ closes the system of differential equations, allowing us to write $\langle ABC \rangle$ in terms of first and second order correlations.

## C. Behaviour in the thermodynamic limit

In Fig. S8 we present the theoretical $N$-dependence of the charging time $\tau$, maximum energy density $E_{\mathrm{max}}$, and maximum power density $P_{\mathrm{max}}$ over a wider range of $N$ than shown in Fig. 3 of the main text. This shows that in addition to the decay-dominated (purple) and coupling-dominated (green) behavior described in the main text, a third region occurs at even larger $N$, which we discuss below.

In the main text we discussed the energetic dynamics around the decay-to-coupling dominated crossover regime, as this was the experimental operating region. Moving deeper into coupling-dominated regime does not necessarily improve the energy storage properties. This is illustrated in Fig. S8(b) which shows the simulated four points, corresponding to the circles in Fig. S8(a). Within the coupling-dominated regime, energy stored within the microcavity rapidly oscillates which is not a desirable feature. This occurs because the light and matter degrees of freedom hybridise to form polaritons with upper and lower branches split by Rabi frequency $\pm g\sqrt{N}$, leading to beating between these modes. These oscillations are not present in the experimentally studied crossover region. In this region, dephasing is strong enough to prevent oscillation in energy, yet weak enough to warrant superextensive charging. Therefore, this is the optimal region to produce a QB. Going deeper into the coupling-dominated regime would only be advantageous if energy was extracted on a shorter timescale than the period of oscillations, or additional mechanisms were in place to stabilise the oscillations.

At even larger $N$ (red region) the stored energy falls with increasing $N$. This can be understood as arising from a condition where the polariton energy splitting exceeds the bandwidth of the pump (set by its finite pulse duration), suppressing energy absorption. Numerically, we find this occurs when $N > N_\sigma$ where $g\sqrt{N_\sigma} = (2/5)^{\frac{1}{4}} (1/\sigma)$, which signifies the onset of this non-resonant regime. The prefactor $(2/5)^{\frac{1}{4}}$ will be explained in Section S2 D. To build an efficient QB in this regime, one should tune the frequency of the laser to match the polariton energies. Additionally, the time dynamics of energy absorption here change significantly, with the second half of the laser pulse causing stimulated emission, reducing the stored energy — such dynamics arises naturally from a toy model of strongly coupled modes with a splitting larger than the pulse bandwidth, and can be seen in the form of the red line in Fig. S8(b).

In Figure S9, we show the theoretical $N$-dependence of $\tau$, $E_{\mathrm{max}}$ and $P_{\mathrm{max}}$ when the frequency of the driving laser is tuned resonant to the lower polariton, i.e. $\Delta_a = \Delta_c = g\sqrt{N}$ in the cumulant equations given in Section S2 B. We emphasise that this is not the condition under which the experiments were performed, but of theoretical interest. By comparison of Figures S9 and S8 one can see that the behaviour of the measures with the different driving frequencies are the same until $N > N_\sigma$. In Figure S9 when $N > N_\sigma$, the laser frequency continues to drive at the frequency of the lower polariton, instead of at the molecular energy as in Figure S8. In this case, the total energy and power in the cavity continue to grow linearly with $N$, and so the energy and power densities are constant.

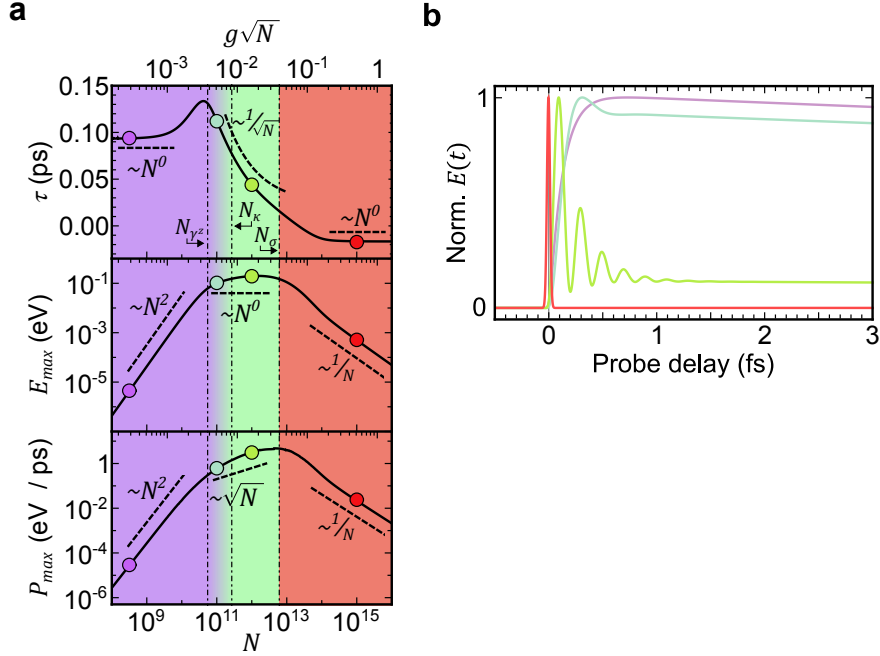

FIG. S8. **Charging dynamics vs number of molecules** $N$. (a) Charging time, peak stored energy, and maximum power as a function of $N$. This figure is identical to Fig. 3(a) in the main text, but extended to larger range of $N$. At large $N$ the stored energy can reach half maximum before the laser pulse finishes, in which case the charging time $\tau$ becomes negative. (b) Examples of dynamics in each regime. The values of $N$ used in these dynamics corresponds to the circles of the same colour in (a).

### D. The boundary between decay and coupling dominant regimes

There are two timescales in this system: the vacuum Rabi splitting (i.e. polariton detuning) $g\sqrt{N}$ and the Rabi splitting $g\sqrt{rN}$ where $rN$ is the number of photons in the cavity. In Fig. 3(a), we show that the microcavity charges super-extensively once $g\sqrt{N}$ is greater than all decay channels. However, this is only true if $r \leq 1$, as is true in experiments A1, A2 and A3. In Fig. 3(b), the boundaries $N_\kappa$ and $N_{\gamma^z}$ are instead determined by $g\sqrt{rN}$ being equal to the decay rates. This is because $r \geq 1$ in experiments B1 and B2. More generally, the important timescale is the larger of the polariton detuning and the Rabi splitting, and so the coupling dominant regime occurs when $g\sqrt{\text{Max}(1, r)N}$ is larger than all decay channels.

In Figure S10 we plot the charging time $\tau$ as a function of $N$ and $r$. Here, we set $\kappa = \gamma^- \equiv \gamma^z \equiv \Gamma = 2$ meV (note that $\gamma^z$ is independent of $N$) so that there is only one boundary between the decay dominant and coupling dominant regimes. The green, red and dashed-black lines show the boundaries between the decay dominant and coupling dominant regimes ($N = N_\Gamma$) if $g\sqrt{N}$, $g\sqrt{rN}$ or $g\sqrt{\text{Max}(1, r)N}$ are used as the relevant coupling scale respectively. Clearly, the boundary is determined by $g\sqrt{\text{Max}(1, r)N}$ for all values of $r$. We also show the boundary between the coupling dominant and non-resonant regimes ($N = N_\sigma$) as the cyan line. When $r > 1$, we find that $N_\sigma$ becomes linearly dependent on $r$. The prefactor $(2/5)^{\frac{1}{4}}$ is necessary for $N_\sigma$ to align with the contours of increased charging time for $r > 1$.

### E. Dependence on laser intensity

Figure S11 shows how capacity, charging time and power vary as a function of laser intensity $r$ at fixed number of molecules $N$. For small $r$, we find that the maximum energy and power densities vary linearly with $r$, while charging time is constant. This simply reflects the total energy in the cavity. The charging time is constant because decay channels still dominate over coherent dynamics. As $r$ is increased beyond $r = 1$, the important timescale $g\sqrt{\text{max}(1, r)N}$ begins to scale with $r$, and so the boundaries separating the coupling dominant and decay dominant regions $N_\kappa$ and $N_{\gamma^z}$ are pushed to smaller $N$. When these boundaries become smaller than the number of molecules in the cavity, the charging time begins to scale as $1/\sqrt{r}$. Additionally, the energy density begins to saturate because there are already many more photons than there are molecules within the cavity. In Figure S11(b) we also plot the experimentally measured energy densities, and we see there is good agreement to the theoretical curve. The coloured points in Figure S11(a) indicate the charging time, maximum capacity and maximum power of the temporal dynamics of the same colour in Figure S11(b).

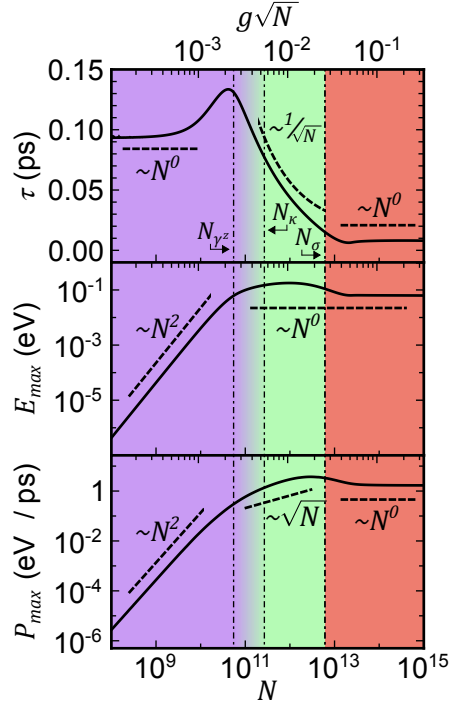

FIG. S9. **Charging dynamics when pumping at lower polariton.** Charging time, peak stored energy, and maximum power as a function of $N$. This figure is identical to Fig. S8 except that the frequency of the laser is tuned to the lower polariton energy, $\Delta_a = \Delta_c = g\sqrt{N}$, rather than the molecular energy.

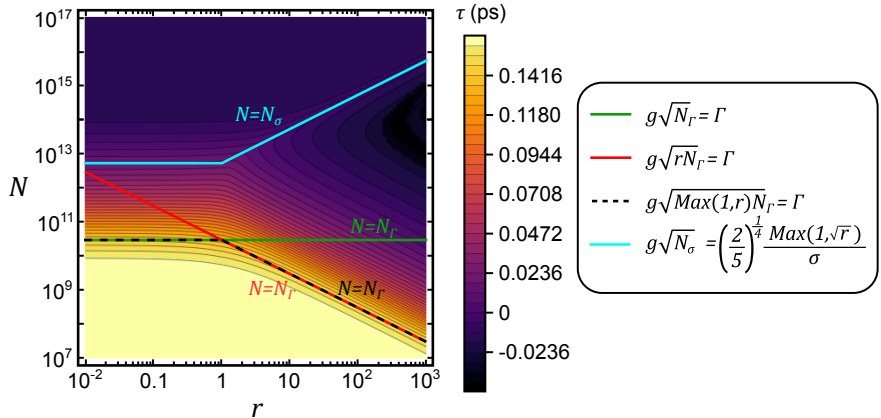

FIG. S10. **Charging time as a function of number of molecules $N$ and laser intensity $r$.** All parameters are equivalent to the Q1% cavity (see main text) with the exception that the dephasing and non-radiative decay rates are equal to the cavity leakage rate (set to $\gamma^z = \kappa = \Gamma = 2$ meV) and note that the dephasing rate is independent of $N$.

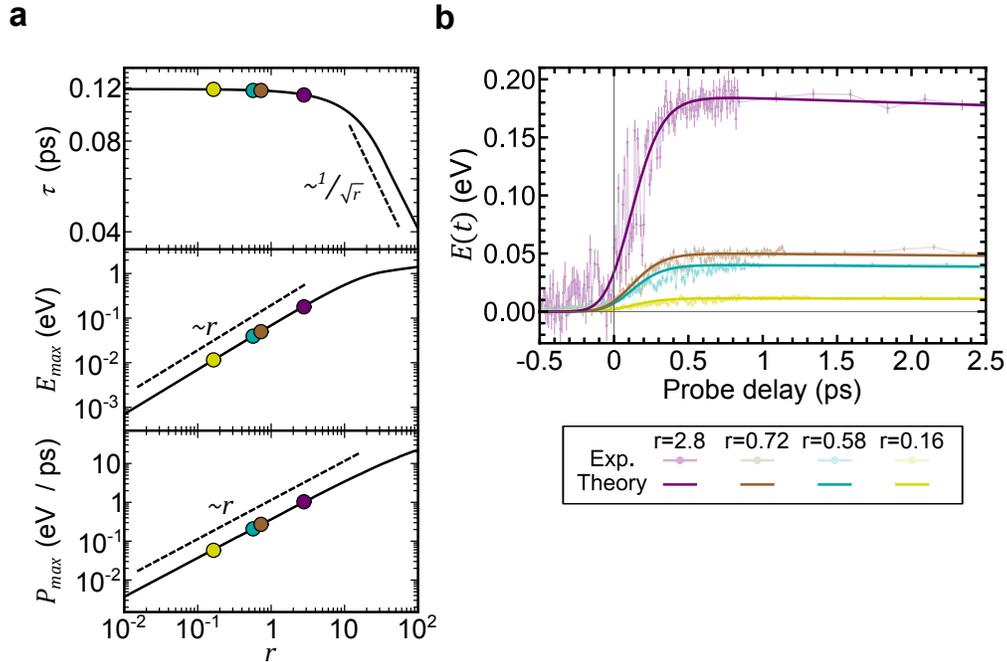

FIG. S11. **Charging dynamics vs pump intensity** $r$. (a) Capacity, charging time and maximum power vs $r$ for a fixed number of molecules $N = 1.62 \times 10^{10}$ (the 1% cavity). All other parameters align with those used to model the 1% cavity. (b) Comparison of theoretical and experimental charging dynamics for four values of $r$. These are indicated by circles in the charging time panel of (a), showing the experimentally measured values.

## S3. FITTING OF MODEL PARAMETERS

As outlined in the main text, we used a reduced chi-squared optimisation procedure to determine the light matter coupling $g$, dephasing constant $\gamma_0^z$ and non-radiative decay rate $\gamma^-$, as well as to estimate uncertainties on these parameters. As these quantities represent molecular properties, we would expect them to be the same in all the different experiments. For this reason, we performed a global fit rather than performing the procedure individually for each experiment. In this section we give further details on this calculation. As also discussed in the main text, our fitting process is key in estimating the charging time, stored energy and peak charging power. Extracting these quantities requires a smooth curve of charge vs time. Any smoothing process implicitly introduces an effective fitting function (such as e.g. fitting the data to piecewise cubic splines). Since our best guess for such a fitting function is in fact the theoretical model described above, we use this continuous function, fit to the experimental data, to extract charging times, energies, and powers.

In performing our fitting, it is necessary to use the values of the molecule numbers, estimated as discussed above. Because of this, errors in the estimates of $N$ affect the fitting parameters, and thus the resulting charging timescales and powers. These correlated errors make it challenging to estimate errors in the power-law scaling of charging time vs $N$.

The cavity lifetime (or equivalently the cavity linewidth $\kappa$) can also be considered as a fitting parameter, however the value of this parameter is more strongly constrained by other measurements. As noted above, transfer matrix simulations on the designed cavity give an upper bound of 306fs. An alternate estimate is provided by comparing the theoretical and measured reflectivity spectra of the cavities. As discussed below, this implies a cavity lifetime of 120fs. In the following, we first present fitting results for a cavity lifetime of 120fs, and then in Sec. S3 B we discuss how the results change with alternate cavity lifetimes in the range 120fs to 306fs.

## A. Fitting procedure

The steps for our fitting procedure are as follows:

1. Calculate the theoretical $E(t)$ curves for a grid of parameter values $g, \gamma^z, \gamma^-$, along with the values of $N$ relevant for all five experiments, A1, A2, A3, B1 and B2. Based on previous observations, we chose the search region of the parameter space as $g \in [0.1, 5000]$ neV; $\gamma_0^z \in [0.1, 5000]$ meV and $\gamma^- = [0.001, 1]$ meV. Subsequent refinements of this search region were made to give higher resolution near the optimal point.

2. We estimate uncertainties, $\sigma_i$ on each experimental data point (transient reflectivity vs time), by considering the point-to-point variation. Because the uncertainty is higher near $t = 0$, when the pump arrives, we use different error estimates in different time windows. Specifically, we divide experiments A1 and A2 into five windows $t < -300\text{fs}; -300\text{fs} < t < 300\text{fs}; 300\text{fs} < t < 700\text{fs}; 700\text{fs} < t < 1000\text{fs}; t > 1000\text{fs}$. For experiments A3, B1 and B2 we found that four windows $t < -300\text{fs}; -300\text{fs} < t < 300\text{fs}; 300\text{fs} < t < 1000\text{fs}; t > 1000\text{fs}$ was sufficient. In each window, the uncertainty estimate for each experiment is taken from the variance over a narrow range of points (typically 150fs) where there is no strong time dependence.

3. For each set of parameters, we performed an "internal" chi-squared minimisation to find the optimal scaling factor $S$ between the stored energy $E(t)$ and the measured differential reflectivity $\Delta R/R$, and a time shift between the theory and experiment $T_0$. That is, we minimise

$$\chi^2 = \sum_i \left[ \frac{S \times (\Delta R/R)_i - E(t_i + T_0)}{\sigma_i} \right]^2, \tag{21}$$

with respect to $S$ and $T_0$. We treat the result of this minimisation as the chi-squared value which we use in the following steps to estimate the meaningful parameters $g, \gamma^z, \gamma^-$ and their uncertainties.

Estimating the scaling factor $S$ from first principles is difficult because of reflections by the cavity mirror, hence this factor is found by the best fit value. The time shift reflects uncertainty of delays in the optics, so that it is not a-prior clear when the peak of the pump pulse arrives. After this shift, we define $t = 0$ as the moment the pump arrives. This is important when calculating the charging time $\tau$, which we defined as the time from the arrival of the pump until reaching half maximum energy.

4. We then use the chi-squared value described above, and divide by the total number of degrees of freedom $k_{\text{eff}} = k - 3$ (where $k$ is the total number of data points), to arrive at the final reduced chi-squared $\tilde{\chi}^2$ map. A slice of this three dimensional reduced chi-squared map for the 120 fs lifetime is shown in Figure S12 for $\gamma^- = 0.0263$ meV, which is the optimal non-radiative decay rate given in the main text. The optimal parameter set used in the main text that optimises $\tilde{\chi}^2$ is shown as the red point in Figure S12. We find $\tilde{\chi}_{min}^2 = 3.048$, suggesting our estimated measurement uncertainties on $\tilde{\chi}$ are reasonable, but likely underestimates.

5. Finally, the 68% confidence interval for each parameter was estimated by considering the contour for which $\tilde{\chi}^2 = \tilde{\chi}_{min}^2 + \frac{1}{k_{\text{eff}}} \Delta^*$ where $\Delta^* = 3.51$ is extracted from the reduced chi-squared distribution for 3 parameters and error tolerance (68%), see [48]. In the right panel of Fig. S12 we show the contour as a white line, and the actual parameter values which lie within this 68% contour as black points.

## B. Fitting cavity lifetime

Figure S13 shows the optimal reduced chi-squared as a function of cavity lifetime, along with the corresponding best-fit values of the parameters $g$, $\gamma_0^z$ and $\gamma^-$, following the fitting procedure described above. The minimal reduced chi-squared is 2.803 occurring for a lifetime of 185 fs.

As noted earlier, the cavity lifetime is also constrained by the measured reflectivity spectrum, shown in Figure S2(d). To check this consistency, Fig. S14 shows the calculated absorption spectra for the 0.5%, 1%, 5% and 10% cavities using the optimal parameter sets given by the 185 fs and 120 fs lifetimes. These are calculated as $\text{Abs}(\Delta\nu) = \text{Re} \int_0^\infty \exp[i(\Delta\nu + \omega)t]\langle a(t)a^\dagger(0)\rangle$ where $\Delta\nu$ is the energy detuning from the cavity and molecules. When evaluated from our model using the quantum regression theorem, we find

$$\text{Abs}(\Delta\nu) = -\text{Re}\left[\frac{i\Delta\nu - \gamma^{\text{tot}}}{\left(i\left[\Delta\nu + \Omega_{\text{eff}}\right] - \frac{2\gamma^{\text{tot}} + \kappa}{4}\right)\left(i\left[\Delta\nu - \Omega_{\text{eff}}\right] - \frac{2\gamma^{\text{tot}} + \kappa}{4}\right)}\right], \tag{22}$$

where $\Omega_{\text{eff}} = \sqrt{g^2 N - (\kappa - 2\gamma^{\text{tot}})^2/4}$ is the effective Rabi splitting. From the measured spectra in Figure S2(d), we expect the 0.5% and 1% cavities to show no polariton splitting, the 5% cavity to have a small splitting, and the 10% cavity to clearly show strong coupling.

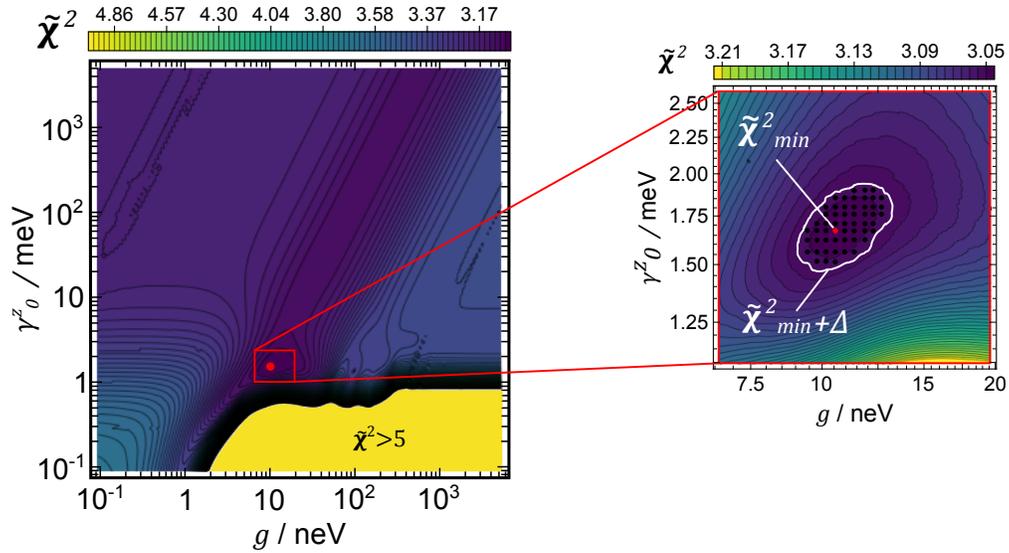

FIG. S12. **Reduced-chi square map to find the optimal parameters for the theoretical model and their 68% confidence intervals.** The chi-squared contour plots shown in this figure are slices of the full three-dimensional map at the optimum non-radiative decay rate $\gamma^- = 0.0141$ meV used in the main text for a cavity lifetime of 120 fs. In the yellow region of the bottom right corner $\tilde{\chi}^2 > 5$, which we do not show to emphasise smaller variations in $\tilde{\chi}^2$. In the right panel, the highlighted contour shows the 68% confidence interval, found using $\Delta = \Delta^*/k_{\text{eff}}$.

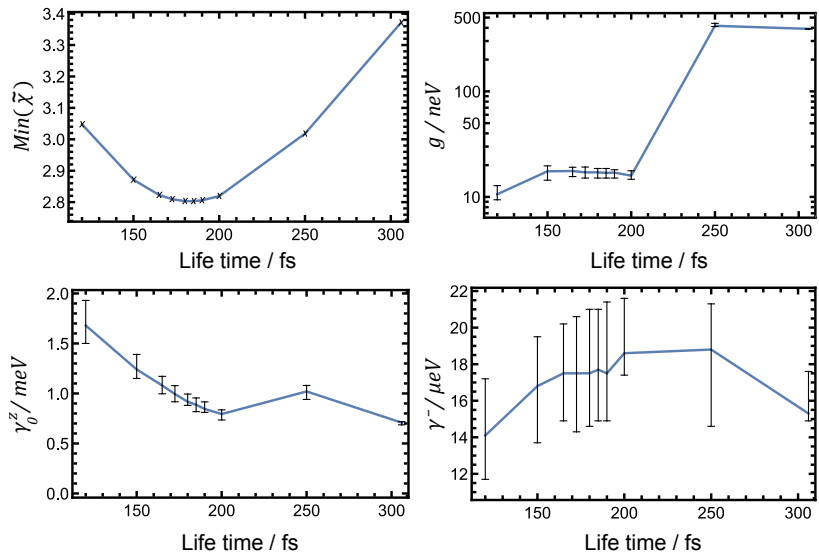

FIG. S13. **Optimal reduced chi-squared, and associated best-fit parameters as a function of cavity lifetime.** Error bars indicate 68% confidence intervals, calculated through the procedure described in Section S3 A.

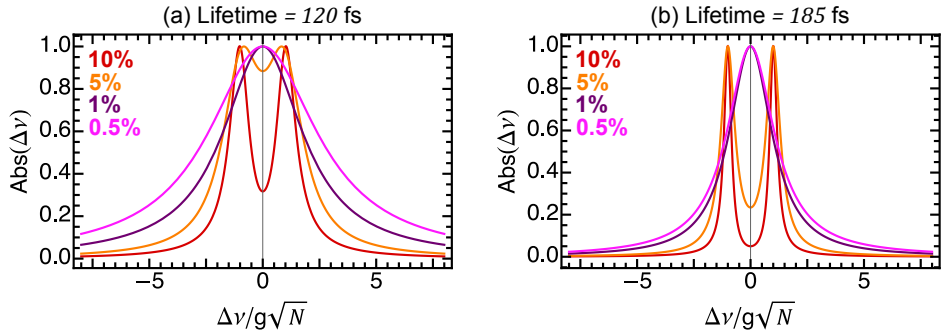

FIG. S14. **Absorption spectra for the** 0.5%, 1%, 5% **and** 10% **cavities**. The spectrum is calculated using Eq. (22) for the best-fit parameters (see Figure S13) for both 120 fs and 185 fs cavity lifetimes. In the 5% and 10% cavities, there are clear polariton peaks forming at $\Delta\nu = \pm g\sqrt{N}$.

It is clear from Figure S14 that although the 185 fs cavity lifetime gives a smaller reduced chi-squared, these parameters predict significantly stronger coupling than is seen in experimental reflectivity spectra. In contrast, the 120 fs cavity lifetime parameters reproduce the polariton splittings across all cavities more accurately, while showing a reduced chi-squared that is not significantly larger. We therefore conclude that 120 fs cavity lifetime, with $g$, $\gamma_0^z$ and $\gamma^-$ given in the main text provides the best fit to the experimental data. In Table S2, we summarise the scaling factors $S$ and time shifts $T_0$ that relate the measured differential reflectivity $\Delta R/R$ to the theoretically calculated stored energy $E(t)$, as used to calculate the fits plotted in Figure 2(b) of the main text. For a formal defintion of $S, T_0$, see Eq. (21).

| Experiment | Scaling factor, $S$ | Time shift, $T_0$/ fs |
|---|---|---|
| **A1** | 2.32 | 47.4 |
| **A2** | 2.01 | -47.4 |
| **A3** | 2.93 | -140.0 |
| **B1** | 3.75 | -159.8 |
| **B2** | 6.24 | -210.5 |

TABLE S2. **The optimal scaling factors and time shifts used to calculate the theoretical curves in Figure 2(b) in the main text.**

### C. Results of fitting procedure

The theoretical time evolution arising from using the above fitting procedure is shown in Figure 3 of the main text. In plotting that figure, the results of the theoretical fit are convolved with an instrument response function, as required to match the experimental data. Figure S15 shows the same data but without convolution by the instrument response, thus providing a more direct picture of the intrinsic dynamics of the system. From the theoretical curves, one can extract the rise time of stored energy $\tau$, the peak stored energy $E_{\max}$, and maximum charging power $P_{\max}$. These values are summarised in Table 1 in the main text.

As well as the scaling to the observables with $N$ shown in Figure 3 in the main text, we could also estimate an effective power-law scaling of the observables $q_i \in \{\tau, E_{\max}, P_{\max}\}$ directly from pairs of experiments $i, j$ by the relation $q_i/q_j = (N_i/N_j)^{f_q}$. As all our observables are intensive (i.e. densities), $f_q > 0$ indicates superextensive behaviour, $f_q = 0$ indicates extensive, and $f_q < 0$ subextensive behaviours. Table S3 gives the observed values of $f_q$.

### D. Residuals of the best fit

To check whether systematic errors arise from our fitting procedure, Fig. S16 shows the residual errors—i.e. difference between the theoretical curves and the experimental data—for the five experiments shown in

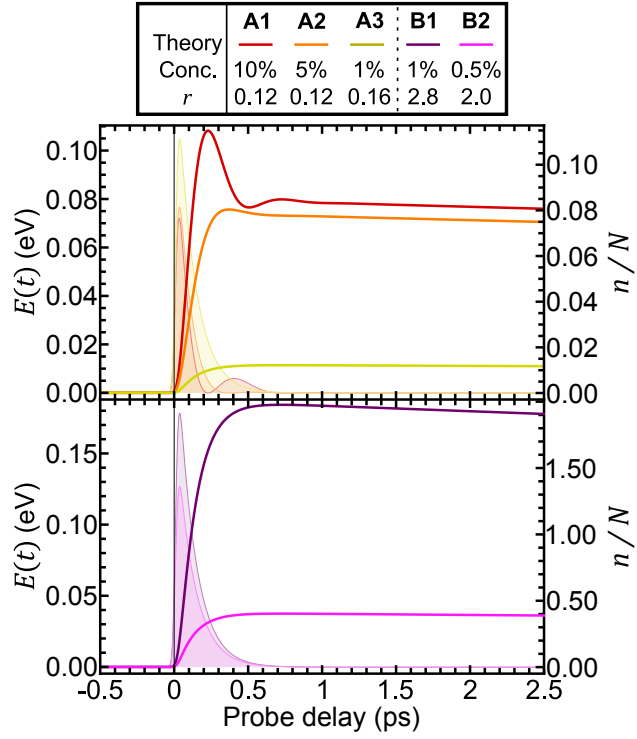

FIG. S15. **Further details of the dynamics of the quantum batteries**. These plots show the time dependence of energy without convolution by the instrument response function (lines without fill) and the ratio of the number of photons, $n$ to molecules, $N$ in the cavity (lines with fill).

| Experiments | $f_\tau$ | $f_{E_{\max}}$ | $f_{P_{\max}}$ |
|---|---|---|---|
| $^{\mathbf{A1}}/_{\mathbf{A2}}$ | -0.35 | 0.52 | 0.94 |
| $^{\mathbf{A2}}/_{\mathbf{A3}}$ | 0.01 | 1.18 | 1.20 |
| $^{\mathbf{B1}}/_{\mathbf{B2}}$ | 0.12 | 2.30 | 2.19 |

TABLE S3. **Observed subextensive and superextensive scaling behaviours in rise-time, stored energy, and charging power.** Power-law exponent $f_q$ for observable $q \in \{\tau, E_{\max}, P_{\max}\}$, where $f_q > 0$, $f_q = 0$, $f_q < 0$ indicates superextensivity, extensivity, and subextensivity, respectively. Table values indicate that charging time $\tau$ is subextensive, whilst stored energy $E_{\max}$ and charging power $P_{\max}$ are superextensive. The first column indicates the corresponding experiments.

Figure 2 of the main text. As is clear, there are no discernible features in the residuals that are consistently present across the different experiments. This indicates that the theoretical curves account for the essential characteristics of the data.

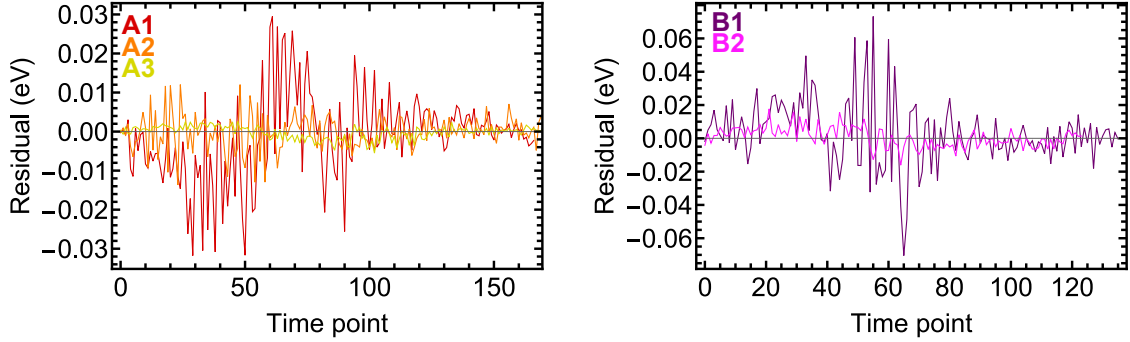

FIG. S16. The residuals of the theoretical curves and the experimental data over the duration of the experiments.

### E. Theoretical fits for 185 fs lifetime

In Figure S17, we show the theoretical fits to the experimental data for experiments A1, A2, A3, B1 and B2 for a lifetime of 185 fs. This lifetime gave the optimal reduce chi-square in Figure S13. From the reduced chi-square fitting procedure, we found that the optimal parameters for this lifetime were $g = 16.9^{+1.7}_{-1.8}$ neV, $\gamma_0^z = 0.887^{+0.068}_{-0.060}$ meV and $\gamma^- = 0.0177^{+0.0033}_{-0.0029}$ meV.

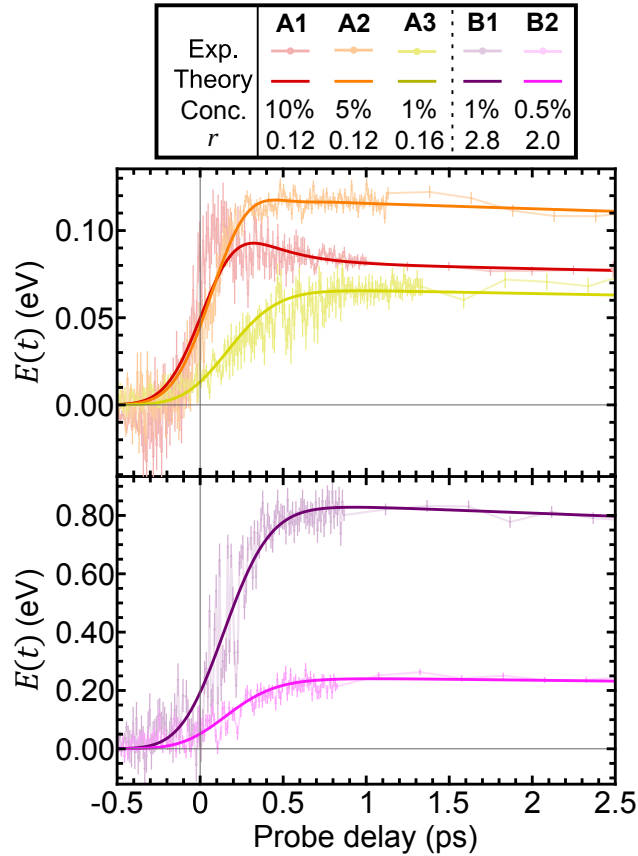

FIG. S17. The fits to the experimental data using a lifetime of 185 fs.